\documentclass[twocolumn,aps,superscriptaddress,floatfix,prb,nobibnotes,longbibliography]{revtex4}
\usepackage[normalem]{ulem}

\usepackage[T1]{fontenc}
\usepackage{graphicx}
\usepackage{amsmath}
\usepackage{amssymb}
\usepackage{bbm}
\usepackage{xspace}
\usepackage{color}
\usepackage{footnote}
\usepackage{comment}
\usepackage{hyperref}
\usepackage{subfigure}
\usepackage{bbold}
\usepackage{ulem}
\usepackage{braket}
\usepackage{multirow}
\usepackage{enumerate}

\definecolor{grey}{RGB}{100,100,100}

\definecolor{orange}{RGB}{252,77,6}
\definecolor{brown}{RGB}{200,127,50}
\definecolor{blue}{RGB}{00,000,100}
\definecolor{blue2}{RGB}{00,000,250}
\definecolor{green1}{RGB}{00,100,00}
\definecolor{green2}{RGB}{00,150,00}
\definecolor{green3}{RGB}{00,200,00}
\definecolor{green4}{RGB}{00,250,00}
\definecolor{dgreen}{rgb}{0.0,0.5,0.0}

\begin{document}

\title{
Hubbard model on the kagome lattice with time-reversal invariant flux and spin-orbit coupling}

\author{Irakli Titvinidze}
\email{titvinidze@itp.uni-frankfurt.de}
\affiliation{Institut f\"ur Theoretische Physik, Goethe-Universit\"at, 60438 Frankfurt am Main, Germany}

\author{Julian Legendre}
\affiliation{CPHT, CNRS, Institut Polytechnique de Paris, Route de Saclay, 91128 Palaiseau, France}

\author{Karyn Le Hur}
\affiliation{CPHT, CNRS, Institut Polytechnique de Paris, Route de Saclay, 91128 Palaiseau, France}

\author{Walter Hofstetter}
\affiliation{Institut f\"ur Theoretische Physik, Goethe-Universit\"at, 60438 Frankfurt am Main, Germany}

\date{\today}

\begin{abstract} 
We study the Hubbard model with time-reversal invariant flux and spin-orbit coupling and position-dependent onsite energies on the kagome lattice, using numerical and analytical methods.  In particular, we perform calculations using real space dynamical mean-field theory (R-DMFT).  
To study the topological properties of the system, we use the topological Hamiltonian approach. 
We obtain a rich phase diagram: for weak and intermediate interactions, depending on the model parameters, the system is in the band insulator, topological insulator, or metallic phase,  while for strong interactions the system is in the Mott insulator phase.
We also investigate the magnetic phases that occur in this system. For this purpose, in addition to R-DMFT, we also use two analytical methods: perturbation theory for large interactions and onsite energies, and stochastic mean-field theory.
\end{abstract}


\maketitle
%

\begin{figure}[t]
\includegraphics[width=8cm]{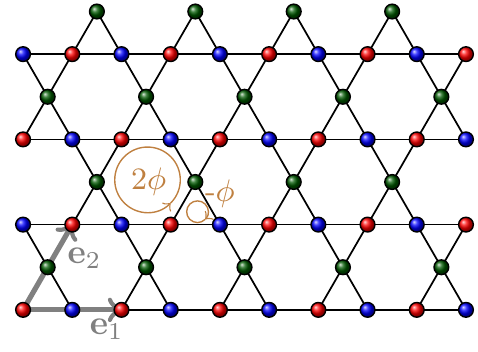}
\caption{
The schematic representation of the kagome lattice. The lattice contains three sites  per unit cell, which we depict by red ($R$), blue ($B$), and green ($G$). ${\bf e}_1$ and ${\bf e}_2$ are the displacement vectors between the neighboring unit cells which form the triangular lattice. The flux is shown here for $\sigma=\uparrow$ fermions.
}
\label{Fig:schematic}
\end{figure}

\section{Introduction}

Ultracold atoms in optical lattices offer new insights into strongly correlated condensed matter\cite{bl.da.08, geor.07, le.sa.07, ho.qi.18}. 
In particular, the fermionic Hubbard model was realized experimentally and the transition from the metallic phase to the Mott insulator phase was observed\cite{ko.mo.05, jo.st.08, sc.ha.08}, as well as the emergence of quantum magnetic order in itinerant systems for large onsite interactions
\cite{du.de.03, si.ba.11, tr.ch.08, ma.ca.16, mu.de.15, ha.du.15, hu.du.16, al.ho.03, ma.ch.17}.
Furthermore, ultracold atoms in optical lattices with synthetic gauge fields can be used to realize topological insulators\cite{ai.at.13, ai.lo.15, mi.si.13,  jo.me.14, fl.re.16, ma.pa.15, st.lu.15}. 
In particular, the Haldane \cite{jo.me.14, fl.re.16} and the Azbel-Harper-Hofstadter \cite{ai.at.13,ai.lo.15,mi.si.13} models were experimentally realized. 
Ultracold atoms in synthetic dimensions also allow to realize robust edge states, which is one of the indicators of topological insulators \cite{ma.pa.15, st.lu.15}.  
%
%
Experimentally, spin-orbit coupling (SOC) has already been studied for ultracold atoms in the absence of optical lattices\cite{li.ji.11, wa.yu.12, ch.so.12,hu.me.16}, 
as well as for one\cite{at.ai.14, li.hu.16} and two\cite{wu.zh.16, su.wa.18} dimensional lattices. There are also other suggestions on how the SOC can be implemented in presence of optical lattices\cite{li.la.14, du.di.04, gr.li.17}.

Ultracold atomic gases in optical lattices allow to realize systems with artificial gauge fields and local Hubbard interaction. Therefore it is of hight interest to study the effect of the local Hubbard interaction on the topological properties of the system. 
In particular, the following aspects have been studied:
the time reversal invariant Hofstadter-Hubbard model\cite{co.or.12, or.co.13, ku.th.16, ir.zh.19, ir.zh.20}, the Haldane-Hubbard model\cite{va.su.11, va.pe.15, yi.hu.21, sh.yu.21}, the Kane-Mele-Hubbard model\cite{ra.le.10, wu.ra.12, pl.va.18, hu.kl.21}, the interacting Rice-Mele model\cite{li.ke.20},  the Bernevig-Hughes-Zhang Hubbard model\cite{am.bu.15, ro.go.16}, Weyl-Hubbard model\cite{ir.gr.21, la.pl.16, ro.go.17}, SU(3) systems with artificial gauge fields\cite{ha.ho.18, ha.zh.19}, and  the Kondo lattice model\cite{we.as.13, hu.zh.18, gr.co.19}.

In materials, the Coulomb interaction can have a very important effect on the topological properties. For instance, in the strongly correlated material $\textrm{Co}_3\textrm{Sn}_2\textrm{S}_2$, for which the Hubbard interaction strength has been estimated around $4$ eV \cite{xu.zh.20}, the magnetic properties have a direct impact on the topological properties \cite{gu.ve.20,le.hu.20, oh.mu.00, xu.li.15}. First principle calculations indicate that the Co (namely the Co-3$d$) orbitals contribute significantly to the electronic properties at the Fermi energy \cite{oh.mu.00, xu.li.15, wa.xu.18,li.su.18,xu.li.18}. Moreover, the magnetic properties of $\textrm{Co}_3\textrm{Sn}_2\textrm{S}_2$, are highly dependent on the Hubbard interactions \cite{xu.zh.20,ir.sk.21,ro.iv.21}. These results therefore point out the very important effect of the Hubbard interactions on the topological properties in $\textrm{Co}_3\textrm{Sn}_2\textrm{S}_2$. This is one of the examples that motivates the study of strong correlations effects in kagome topological systems.

Here we study the Hubbard model, with time-reversal invariant flux and with spin-orbit coupling and position-dependent onsite energies on the kagome lattice, which is a non-Bravais lattice and contains three sites per unit cell.
Experimentally, this lattice has been realized using ultracold atoms by superimposing two triangular optical lattices with different wavelengths\cite{jo.gu.12}. 
We have studied the same model, but without interaction, in Ref.~\onlinecite{ti.le.21}, where depending on the model parameters, we have obtained a topological insulator, band insulator, or metallic phase. 
There are a number of other works that have already investigated topological  properties of the non-interacting tight-binding model on the kagome lattice\cite{oh.mu.00, ko.ho.10, gr.sa.10, zhan.11, pe.ho.12, xu.li.15, li.su.18, gu.ve.20, le.hu.20, gu.fr.09, li.zh.09,  wa.zh.10, li.zh.10, ta.me.11,  li.ch.12, li.ch.13, ch.ch.14, du.ch.18, bo.na.19, ku.yo.19, wa.we.19}.
On the other hand, the Hubbard model on the kagome lattice without the flux has also been intensively studied\cite{miel.92, oh.ka.06, oh.su.07, fu.oh.10,ya.se.11, ya.se.12a, ya.se.12b, ka.st.21,
be.ca.07, ku.ka.07,   ki.oh.13, hi.as.16, su.zh.21, ud.mo.10, fe.ra.14, kr.th.12, ki.pl.13, guer.14}, 
in particular,  the metal insulator transition\cite{miel.92, oh.ka.06, be.ca.07, ku.ka.07, oh.su.07, fu.oh.10, ya.se.11, ya.se.12a, ya.se.12b, ki.oh.13, hi.as.16, ka.st.21, su.zh.21} and magnetic order\cite{oh.ka.06, oh.su.07, fu.oh.10,ya.se.11, ya.se.12a, ya.se.12b, ka.st.21, ki.za.15, wa.li.16}.

To investigate this problem, we use numerical and analytical methods.  One of the most powerful methods for studying strongly correlated systems in two and higher dimensions is dynamical mean-field theory (DMFT)\cite{ge.ko.96, me.vo.89}. Here we use its real space generalization, real-space DMFT (R-DMFT) \cite{sn.ti.08, he.co.08, po.no.99, co.or.12}. 
To study topological properties of the system we use the topological Hamiltonian approach\cite{wa.zh.12, ku.me.16}. 
To study the magnetic properties of the system for large interactions and onsite energies, we also apply perturbation theory and stochastic mean-field theory\cite{hu.kl.21}.

The paper is organized as follows. In the next section (Sec.~\ref{Model}), we introduce the model Hamiltonian.
In Sec. ~\ref{Method}, we give an overview of the applied methods. In particular, in Sec. ~\ref{RDMFT-Topological_Hamiltonian} we give an overview of R-DMFT and the topological Hamiltonian approach, while our analytical methods are presented in Secs. ~\ref{Perturbation_lambdaR_U} and ~\ref{Stochastic_MF}. 
Afterwards, we present our results in sections \ref{Results_without} and \ref{section_staggered_potential}. 
In section ~\ref{Results_without} we consider onsite energies applied only to ``red'' sublattice sites (see Fig. \ref{Fig:schematic}), while in section ~\ref{section_staggered_potential} we consider a setup with a staggered potential. Finally, concluding remarks are made in Sec.~\ref{Conclusions}.

\section{Model}\label{Model}

We study the 
fermionic Hubbard model with time-reversal invariant flux and Rashba-type  spin-orbit coupling on the kagome lattice, which is a non-Bravais lattice and has three sites per unit cell. The unit cells are arranged on a triangular lattice. 
The Hamiltonian has the form 
\begin{subequations}
\label{Hamiltonian}
\begin{align}
\label{Hamiltonian_tot}
{\cal H}
&={\cal H}_{0}+{\cal H}_{U} \\
\label{Hamiltonian0}
{\cal H}_{0}
&=-t\sum_{{\bf r}}\Biggl[
c_{{\bf r},R}^{\dagger}\mathbb{1}c_{{\bf r},B}^{\phantom\dagger}
+c_{{\bf r}+{\bf e}_1,R}^{\dagger}\mathbb{1}c_{{\bf r},B}^{\phantom\dagger} 
\nonumber
\\
&+c_{{\bf r},R}^{\dagger}e^{-i2\pi\gamma\sigma^x}c_{{\bf r},G}^{\phantom\dagger}
+c_{{\bf r},G}^{\dagger}e^{-i2\pi\gamma\sigma^x}c_{{\bf r}+{\bf e}_2,R}^{\phantom\dagger}
\nonumber\\
&+c_{{\bf r},B}^{\dagger}e^{i\phi \sigma^z}c_{{\bf r},G}^{\phantom\dagger}+c_{{\bf r}+{\bf e}_3,B}^{\dagger}e^{i\phi\sigma^z}c_{{\bf r},G}^{\phantom\dagger} 
+ H.c.\Biggl] 
\nonumber \\
&+\sum_{{\bf r}}\sum_{\alpha=R,B,G} (V_{{\bf r},\alpha} -\mu)n_{{\bf r},\alpha} 
\\
\label{HamiltonianU}
{\cal H}_{U}&=
U\sum_{\bf r}\sum_{\alpha=R,B,G} n_{{\bf r},\alpha,\uparrow}n_{{\bf r},\alpha,\downarrow} \, .
\end{align}
\end{subequations}
Here $c_{\alpha,{\bf r}}^{\dagger}=\left(c_{\alpha,{\bf r},\uparrow}^{\dagger},c_{\alpha, {\bf r},\downarrow}^{\dagger}\right)$ is the fermionic creation operator at site $\{{\bf r}, \alpha\}$, where $\bf r$ is a coordinate of the unit cell and $\alpha=R,G,B$ specifies the location within the unit cell (see Fig.~\ref{Fig:schematic}).  $n_{{\bf r},\alpha,\sigma}=c_{\alpha,{\bf r},\sigma}^{\dagger},c_{\alpha, {\bf r},\sigma}^{\phantom\dagger} $ is the fermion number operator for spin $\sigma$ on the corresponding site and $n_{{\bf r},\alpha}=n_{{\bf r},\alpha,\uparrow}+n_{{\bf r},\alpha,\downarrow}$. 
${\bf e}_1$ and ${\bf e}_2$ are the basis vectors of the triangular lattice. 
$\sigma^x$ and $\sigma^z$ are the $x$ and $z$ Pauli matrices acting in spin space, while $\mathbb{1}$ is the identity matrix. 
$t$ is the hopping amplitude of the fermions between neighboring lattice sites and is taken as the energy unit.
We include Rashba-type spin-orbit coupling with strength $\gamma$ for the hopping between neighboring $R$ and $G$ sublattice sites 
(second line in Eq. \eqref{Hamiltonian0}).
When fermions hop from the $G$ to the $B$ sublattice site, they acquire  the phase $s_z\phi$, which has opposite sign for $\sigma=\uparrow$ ($s_z=1$) and $\sigma=\downarrow$ ($s_z=-1$) particles, so that the Hamiltonian preserves time-reversal symmetry. 
$V_{{\bf r},\alpha}$ is the onsite energy on the lattice site $\{{\bf r},\alpha\}$. 
Finally $\mu$ is the chemical potential and $U$ is the local Hubbard interaction. 
We define the filling  ${n=\frac{1}{3N_1N_2}\sum\limits_{{\bf r},\alpha,\sigma}\langle n_{{\bf r},\alpha,\sigma}\rangle}$.
$N_1$ and $N_2$ are the numbers of unit cells of the kagome lattice along the directions ${\bf e}_1$ and ${\bf e}_2$.

In our previous work\cite{ti.le.21}, we studied the same model, but without local Hubbard interaction ($U=0$). 
We obtain that without staggered potential gaps may appear between the second and third bands, when two from six bands are filled, as well as between the fourth and fifth bands when four from six bands are filled. Therefore, a gap may appear for the fillings $n=2/3$ and $n=4/3$. Just reminder when all bands are filled $n=2$. In case of staggered potential gaps may appear for the fillings $n=1/3$, $n=2/3$, $n=1$, $n=4/3$, and $n=5/3$.
We considered different arrangements and obtained rich phase diagrams. In particular, we have obtained metallic phases, band and topological insulator phases depending on the model parameters. The topological insulator behaves as insulators in the bulk, while they are conducting at their boundary. Despite that, their topological properties are classified according to the topological invariants which are determined based on their bulk properties. Time-reversal symmetric nonmagnetic insulators are characterized by a ${\mathbb Z}_2$ invariant $\nu$, i.e., they are divided into two categories: a topological insulator with ${\mathbb Z}_2$ number $\nu=1$ and a trivial band insulator with $\nu=0$. The latter is adiabatically connected to the trivial state, while the former cannot be connected to the trivial state without closing a bulk gap\cite{ka.me.05a, ka.me.05b}.

Here we consider two different cases. First, we consider the case where the onsite energy is independent of $\bf r$ and is nonzero only on the $R$ sublattice site, i.e., $V_{{\bf r},R}=\lambda$ and $V_{{\bf r},B}=V_{{\bf r},G}=0$ and we considered $2/3$ filled system.
Next, we consider a staggered potential.  For the latter we have $V_{{\bf r},\alpha}=\lambda$ for ${{\bf r}=2n_1{\bf e}_1+n_2{\bf e}_2}$, and  $V_{{\bf r},\alpha}=-\lambda$ for ${{\bf r}=(2n_1+1){\bf e}_1+n_2{\bf e}_2}$. Here $-\frac{N_1}{4} < n_1 \leq \frac{N_1}{4} $ and $-\frac{N_2}{2} < n_2 \leq \frac{N_2}{2} $ are integer numbers. 
Due to this staggered potential the size of the unit cell of the model is twice as large as the size of the unit cell of the lattice. In this case we considered half-filled system.

\section{Methods}
\label{Method}

\subsection{Real-space dynamical mean-field theory and effective topological Hamiltonian}
\label{RDMFT-Topological_Hamiltonian}

One of the methods we use to study the behavior of the system is real space dynamical mean-field theory (R-DMFT)\cite{sn.ti.08, he.co.08, po.no.99, co.or.12}, a real space generalization of dynamical mean field theory (DMFT)\cite{ge.ko.96, me.vo.89}, which is a powerful tool for studying strongly correlated systems in two and higher dimensions.
It maps the original model with local Hubbard interaction to a set of coupled, self-consistent single-impurity Anderson models (SIAMs)
\begin{align}
{\cal H}_{\rm SIAM}^{(\iota)}&=
Un_{\iota,\uparrow}n_{\iota,\downarrow} +(V_{\iota}-\mu) n_{\iota}
+\sum_{{\bf k},\sigma,\sigma'}\varepsilon_{{\bf k},\sigma\sigma'}^{(\iota)}a_{{\bf k},\sigma}^{\dagger}a_{{\bf k},\sigma}^{\phantom\dagger}
\nonumber\\
&+\sum_{{\bf k},\sigma}\left(v_{{\bf k},\sigma}^{(\iota)} a_{{\bf k},\sigma}^{\dagger} c_{\iota,\sigma}^{\phantom\dagger}+w_{{\bf k},\sigma}^{(\iota)} a_{{\bf k},\sigma}^{\dagger} c_{\iota,\bar\sigma}^{\phantom\dagger}+H.c.\right)
,
\end{align}
one for each lattice site ($\iota=({\bf r},\alpha)$), which has to be solved self-consistently.  Here $a_{{\bf k},\sigma}^{\dagger}$ is creation operator of the non-interacting fermion in the bath with momentum $\bf k$ and spin $\sigma$. $\varepsilon_{{\bf k},\sigma\sigma'}^{(\iota)}$ is a dispersion for $\iota$ site's bath, while $v_{{\bf k},\sigma}^{(\iota)} $ and $w_{{\bf k},\sigma}^{(\iota)}$ describe the spin-conserving and spin-mixing coupling between bath and impurity, respectively. The bar notation applies as $\bar \uparrow =\downarrow$ and $\bar\downarrow=\uparrow$.
We solve the SIAM using exact diagonalization (ED)\cite{ca.kr.94}. In our calculations we used four bath sites.
R-DMFT is non-perturbative in the Hubbard interaction and fully takes into account local quantum fluctuations, but neglects non-local quantum fluctuations. According to R-DMFT, the self-energy is a local quantity but can be position-dependent, i.e.
${\Sigma_{{\bf r},\alpha;{\bf r}',\alpha'}^{\sigma\sigma'}(i\omega_n)=\Sigma_{{\bf r},\alpha}^{\sigma\sigma'}(i\omega_n)\delta_{{\bf r}{\bf r}'}\delta_{\alpha\alpha'}}$. 
Here $\omega_n=(2n+1)\pi T$ is the fermionic Matsubara frequency and $T$ is temperature.

The main idea how R-DMFT works is the following: we start with an initial guess of the self-energy $\Sigma_{{\bf r},\alpha}^{\sigma\sigma'}(i\omega_n)$ at each lattice site. We then calculate the Green's function of the lattice
\begin{equation}
 \label{Lattice_Green} 
G_{{\bf r},\alpha;{\bf r}',\alpha'}^{\sigma\sigma'}(i\omega_n)
= \left[i\omega_n \mathbb{1} - {\cal H}_{0} - 
\Sigma_{{\bf r},\alpha;{\bf r}',\alpha'}^{\sigma\sigma'}(i\omega_n) \right]^{-1} \,,
\end{equation}
which is a $3N_1N_2 \times 3N_1N_2$ matrix. 
The diagonal elements (real space) of the lattice Green's function are identified with the local Green's functions on the different lattice sites. With the knowledge of the local (impurity) Green's functions we determine the Weiss Green's functions
\begin{equation}
\label{Weiss_Green}
{{\cal G}_0}_{{\bf r},\alpha}^{\sigma\sigma'}(i\omega_n)
=\left[[G_{{\bf r},\alpha;{\bf r},\alpha}^{\sigma\sigma'}(i\omega_n)]^{-1} + \Sigma_{{\bf r},\alpha}^{\sigma\sigma'}(i\omega_n)
\right]^{-1}
\end{equation}
After obtaining the Weiss Green's functions, we determine the model parameters for the SIAMs based on the following expressions for Weiss Green's function
\begin{subequations}
\begin{align}
{{{\cal G}_0}_{{\bf r},\alpha}^{-1}}^{\sigma\sigma}(i\omega_n) 
&= i\omega_n + \mu - V_{{\bf r},\alpha}+ \sum_{\bf k}\frac{|v_{{\bf k},\sigma}^{(\iota)}|^2 + |w_{{\bf k},\sigma}^{(\iota)}|^2}{i\omega_n-\varepsilon_{{\bf k},\sigma\sigma}^{(\iota)}}
\\
{{{\cal G}_0}_{{\bf r},\alpha}^{-1}}^{\sigma\bar\sigma}(i\omega_n)
&=-\sum_{\bf k}\frac{v_{{\bf k},\sigma}^{(\iota)}{w_{{\bf k},\sigma}^{(\iota)}}^*+w_{{\bf k},\sigma}^{(\iota)}{v_{{\bf k},\sigma}^{(\iota)}}^*}{i\omega_n-\varepsilon_{{\bf k},\sigma\sigma}^{(\iota)}}
\end{align}
\end{subequations}
Then we solve the SIAMs and obtain new values for the self-energies. We repeat this process until convergence is achieved.  
At this point it is worth mentioning that the obtained results are independent of the initial guess of the self-energies, unless there is a hysteresis region. In the latter case, there can be multiple classes of initial self-energies, and depending on from which we start, we obtain different solutions.

In this work we consider a system containing $N_1 \times N_2=20\times20$ unit cells, i.e. $1200$ lattice sites. We consider periodic boundary conditions (PBC). 
To check the precision of our calculations, we also perform calculations for a system with $N_1 \times N_2=10\times10$ unit cells. We obtain good agreement, which makes us confident that our results for $N_1 \times N_2=20\times20$ unit cells are reliable.
When onsite energies are independent of $\bf r$ and non-zero only on 
$R$ sublattice sites, i.e. $V_{{\bf r},R}=\lambda$ and $V_{{\bf r},B}=V_{{\bf r},G}=0$, the Hamiltonian in Eq. \eqref{Hamiltonian} is symmetric under the translations ${\bf r} \rightarrow {\bf r} + {\bf e}_{1}$ and ${\bf r} \rightarrow {\bf r} + {\bf e}_{2}$.   
Due to this symmetry of the model we consider $3$ distinguishable self-energies  
${\Sigma_{{\bf r},\alpha}^{\sigma\sigma'}(i\omega_n)=\Sigma_{\alpha}^{\sigma\sigma'}(i\omega_n)}$ (${\alpha=R,B,G}$).
On the other hand, in the case of the staggered potential discussed in Sec. \ref{Model} the Hamiltonian in Eq. \eqref{Hamiltonian} is symmetric under the translations ${\bf r} \rightarrow {\bf r} + 2{\bf e}_{1}$ and ${\bf r} \rightarrow {\bf r} + {\bf e}_{2}$.   In this case we consider  $6$ distinguishable self-energies inside the unit cell of the model: $\Sigma_{{\bf r}, \alpha}^{\sigma\sigma'}(i\omega_n)=\Sigma_{\alpha,1}^{\sigma\sigma'}(i\omega_n)$ for ${{\bf r}=2n_1{\bf e}_1+n_2{\bf e}_2}$, and  
$\Sigma_{{\bf r},\alpha}^{\sigma\sigma'}(i\omega_n)=\Sigma_{\alpha,2}^{\sigma\sigma'}(i\omega_n)$ for ${{\bf r}=(2n_1+1){\bf e}_1+n_2{\bf e}_2}$.

Our R-DMFT calculations focus mainly on the paramagnetic solution, which is sufficient to describe the Mott transition. However, we will consider magnetic solutions as well. 
The paramagnetic solution requires that the off-diagonal elements of the self-energy vanish in spin space, while the diagonal elements are equal to each other, i.e.
\begin{subequations}
\begin{align}
&\Sigma_{{\bf r},\alpha}^{\uparrow\uparrow}(i\omega_n)
=\Sigma_{{\bf r},\alpha}^{\downarrow\downarrow}(i\omega_n)
=\Sigma_{{\bf r},\alpha}(i\omega_n)
\\
&\Sigma_{{\bf r},\alpha}^{\uparrow\downarrow}(i\omega_n)
=\Sigma_{{\bf r},\alpha}^{\downarrow\uparrow}(i\omega_n)
=0
\end{align}
\end{subequations}

To detect the transition to the Mott insulator phase we compute the following quantities: the double occupancy 
\begin{equation}
\label{double_occupancy}
d_{{\bf r},\alpha}=\langle n_{{\bf r},\alpha,\uparrow}n_{{\bf r},\alpha,\downarrow}\rangle \,,
\end{equation}
where $\langle \ldots \rangle$ denotes the expectation value and is determined based on solving SIAMs; and the quasi-particle weight 
\begin{equation}
Z_{{\bf r},\alpha}
=\frac{1}{1-\frac{\partial \Sigma_{{\bf r},\alpha}(\omega)}{\partial\omega}} 
=\frac{1}{1-\frac{\Im m\Sigma_{{\bf r},\alpha}(i\omega_0)}{\omega_0}} 
\,.
\end{equation}
In the Mott insulator phase both of these  quantities are strongly suppressed 
$d_{{\bf r},\alpha} \ll 1$ and $Z_{{\bf r},\alpha} \ll 1$.

The topological properties of the interacting system can be determined based on the knowledge of the single particle Green's function $G_{{\bf r},\alpha;{\bf r}',\alpha'}^{\sigma\sigma'}(i\omega_n)$.\cite{is.ma.86, wa.qi.10, gura.11, ku.me.16}  
It can be shown that it is often not necessary to know the Green's function in the entire frequency range, but only the mode $\omega=0$ is crucial\cite{wa.zh.12}.
This method is called the topological Hamiltonian approach\cite{wa.zh.12, ku.me.16} and is a powerful method to compute the Chern number or $\mathbb{Z}_2$ invariant for systems with many-body interactions. The topological Hamiltonian can be written as
\begin{align}
\label{Topological_Hamiltonian} 
{\cal H}_{T}&=-[G_{{\bf r},\alpha;{\bf r}',\alpha'}^{\sigma\sigma'} (i\omega_n\rightarrow 0)]^{-1}
={\cal H}_{0}+\Sigma_{{\bf r},\alpha;{\bf r}',\alpha'}^{\sigma\sigma'}(i\omega_n \rightarrow 0)
\nonumber\\
&={\cal H}_{0}+\Sigma_{{\bf r},\alpha}^{\sigma\sigma'}(i\omega_n \rightarrow 0)\delta_{{\bf r}{\bf r}'}\delta_{\alpha\alpha'} \,.
\end{align}
In this step, we map our original Hamiltonian with many-body interaction to an effective non-interacting Hamiltonian, where the effect of the interaction is included via the self-energies  $\Sigma_{{\bf r},\alpha}^{\sigma\sigma'}(i\omega_n \rightarrow 0)\delta_{{\bf r}{\bf r}'}\delta_{\alpha\alpha'}$, which are determined using R-DMFT. 

Further, we use the above effective non-interacting Hamiltonian ${\cal H}_{T}$ to calculate the $\mathbb{Z}_2$ invariant using the approach employing twisted boundary conditions\cite{fu.ha.05, fu.ha.07, ku.me.16, ir.zh.20, ti.le.21}. 
We consider spin-dependent twisted boundary conditions along the ${\bf e}_1$ direction and spin-independent twisted boundary conditions along the ${\bf e}_2$ direction. So we have
\begin{equation}
\label{twsited_boundary_condition}
c_{{\bf r}+L_1{\bf e}_1,\alpha}=e^{i\sigma_z \theta_{1}}c_{{\bf r},\alpha}
\quad{\rm and}\quad
c_{{\bf r}+L_2{\bf e}_2,\alpha}=e^{i \mathbb{1}\theta_{2}}c_{{\bf r},\alpha}\,.
\end{equation}
Here, $L_1$ and $L_2$ are linear dimensions of the 2D sample area considered for the topological Hamiltonian ${\cal H}_T$. They are in general different from $N_1$ and $N_2$, the system dimensions considered in the R-DMFT calculations.  We perform calculations for a relatively small real-space sample $L_{\kappa=1,2} \leq 8$. We find that the obtained results are independent of the value of $L_\kappa$.
$\boldsymbol{\theta}=(\theta_1,\theta_2)$  is the vector of the two twist angles. $\theta_{\kappa=1,2}=2\pi n_{\kappa}/N_{\theta_\kappa}$, where $-N_{\theta_\kappa}/2 \leq n_{\kappa} <N_{\theta_\kappa}/2$. In our calculations we consider  $N_{\theta_\kappa}=40$.

For time-reversal invariant systems the $\mathbb{Z}_2$ number is then given as\cite{fu.ha.05, fu.ha.07, ku.me.16, ir.zh.20, ti.le.21} 
\begin{equation}
\label{nu_number_definition} 
\nu \equiv \left[\frac{1}{4\pi i} \sum_{\boldsymbol{\theta}}\Omega(\boldsymbol{\theta})\right]  \mod \,\, 2 \,. 
\end{equation}
Here 
\begin{equation}
\label{Bery_curvature}
\Omega(\boldsymbol{\theta})=\log \left[U_1(\boldsymbol{\theta})
U_2(\boldsymbol{\theta}+\boldsymbol{\mu}_{1})
U_1(\boldsymbol{\theta}+\boldsymbol{\mu}_{2})^{-1}
U_2(\boldsymbol{\theta})^{-1}
\right] 
\end{equation}
is the Berry curvature and
\begin{equation}
U_{\kappa=1,2}(\boldsymbol{\theta})=
\frac{\det \langle \psi_a(\boldsymbol{\theta})|\psi_b(\boldsymbol{\theta}+\boldsymbol{\mu}_\kappa)\rangle }{|\det \langle \psi_a(\boldsymbol{\theta})|\psi_b(\boldsymbol{\theta}+\boldsymbol{\mu}_\kappa)\rangle |}
\end{equation}
is  the $U(1)$ link variable which is a function of the twist angle $\boldsymbol{\theta}$. Here ${|\psi_{a(b)}(\boldsymbol{\theta})\rangle}$ are the 
occupied eigenstates of the Hamiltonian for a given twist angle $\boldsymbol{\theta}$.   $\boldsymbol{\mu}_{1}=\left(2\pi/N_{\theta_1},0\right)$ and  $\boldsymbol{\mu}_{2}=\left(0,2\pi/N_{\theta_2}\right)$ 
are unit vectors in the respective directions.

\subsection{Perturbation theory at $\lambda \gg t$ and $U \gg t$}
\label{Perturbation_lambdaR_U}

\subsubsection{Effective Hamiltonian}

Here we consider the Hamiltonian 
\begin{equation}
{\cal H} ={\cal H}_U + {\cal H}_\lambda + {\cal H}_t,
\end{equation}
where ${\cal H}_U$ is defined in Eq.~\eqref{HamiltonianU} and 
\begin{subequations}
\begin{align}
{\cal H}_{t}
&=-t\sum_{{\bf r}}\Biggl[
c_{{\bf r},R}^{\dagger}\mathbb{1}c_{{\bf r},B}^{\phantom\dagger}
+c_{{\bf r}+{\bf e}_1,R}^{\dagger}\mathbb{1}c_{{\bf r},B}^{\phantom\dagger} 
\nonumber
\\
&+c_{{\bf r},R}^{\dagger}e^{-i2\pi\gamma\sigma^x}c_{{\bf r},G}^{\phantom\dagger}
+c_{{\bf r},G}^{\dagger}e^{-i2\pi\gamma\sigma^x}c_{{\bf r}+{\bf e}_2,R}^{\phantom\dagger}
\nonumber\\
&+c_{{\bf r},B}^{\dagger}e^{i\phi \sigma^z}c_{{\bf r},G}^{\phantom\dagger}+c_{{\bf r}+{\bf e}_3,B}^{\dagger}e^{i\phi\sigma^z}c_{{\bf r},G}^{\phantom\dagger} 
+ H.c.\Biggl] \, , \\
{\cal H}_\lambda &= \sum_{{\bf r},\sigma} \lambda \, n_{{\bf r},R,\sigma} \, ,
\end{align}
\end{subequations}
such that ${\cal H}_0 = {\cal H}_t + {\cal H}_\lambda$.
We denote by P an orthogonal and hermitian projector which commutes with ${\cal H}_U + {\cal H}_\lambda$ and $Q=1-P$ the associated complementary projector. We write $G(z)=\left( z - {\cal H}\right)^{-1}$ the resolvent of the Hamiltonian ${\cal H}$. Considering $z=E + i \eta$ or $z=E - i \eta$ gives the Fourier transform (up to a factor $i \hbar$) of respectively the retarded or advanced Green function, which dictates the time evolution under ${\cal H}$. We have
\begin{equation}
PG(z)P=P\left( zP - {\cal H}_{\textrm{eff}}\right)^{-1},
\end{equation}
with
\begin{equation} \label{res}
{\cal H}_{\textrm{eff}} = P {\cal H} P + P {\cal H}_t Q \left( z Q - Q {\cal H} Q\right)^{-1} Q  {\cal H}_t P
\end{equation}
being the effective Hamiltonian for the evolution of the system in the subspace in which $P$ projects.

\subsubsection{$\lambda \gg t$ and $U \gg t$}

We consider $\lambda \gg t$ and $U \gg t$. 
The reason for our interest in this limit is that we want to study the magnetic properties of the system. For this, we need a well-defined spin.
Because of the large onsite energies ($\lambda \gg t$), the system can be described by an effective model of a half-filled square lattice, as we have shown in our earlier work\cite{ti.le.21}.   The large local Hubbard interaction ensures that each side is singly occupied, and accordingly we have a well-defined spin.

We denote by $S_0$  the subspace of eigenvectors of ${\cal H}_U + {\cal H}_\lambda$, corresponding to the ground state with energy $E_0$. 
To second order in $t/U$ and $t/\lambda$, the effective Hamiltonian in $S_0$ is
\begin{align} \label{res2}
{\cal H}_{\textrm{eff}}= &E_0 P + P{\cal H}_tP \nonumber \\ &+  P{\cal H}_t Q \left(E_0 - Q \left( {\cal H}_U + {\cal H}_\lambda \right)Q \right)^{-1} Q{\cal H}_tP
\end{align}
where $P$ here projects on $S_0$. The higher orders can be computed from Eq.~\eqref{res}.
In practice, we consider two states $\ket{\psi}$, and $\ket{\psi '}$ in $S_0$, and from Eq.~\eqref{res2} we get (still at second order in $t/U$ and $t/\lambda$)
\begin{align}
\bra{\psi'} {{\cal H}_{\textrm{eff}}}\ket{\psi}= &E_0 \braket{\psi'|\psi} + \bra{\psi'} {{\cal H}_t}\ket{\psi}  \nonumber \\ &+ \sum_{\ket{m}\notin S_0} \dfrac{\bra{\psi'} {{\cal H}_t} \ket{m} \bra{m} {{\cal H}_t} \ket{\psi}}{E_0-E_m},
\end{align}
with $\ket{m}$ the eigenstates of ${\cal H}_U + {\cal H}_\lambda$ which do not belong to the ground state subspace $S_0$ and $E_m$ is the corresponding eigenenergy.

We implement this method for our model and derive an effective Hamiltonian for $\lambda \gg t$ and $U \gg t$ in section ~\ref{results_for_large_lambda_and_U}.

\subsection{Stochastic mean field method}
\label{Stochastic_MF}

Here we rely on the approach developed in Ref.~\onlinecite{hu.kl.21}. We use a mean-field approximation to rewrite the Hamiltonian. 
Then we search for the value of the mean field parameter that minimizes the associated energy. 
First we rewrite the Hubbard interaction term
\begin{equation}
{\cal H}_U = \dfrac{U}{2} \sum_{{\bf r}} \sum_{\alpha=R,B,G} \left( {\bf S}_{\alpha,{\bf r}} \cdot {\bf S}_{\alpha,{\bf r}} + S_{\alpha,{\bf r}}^{0}  \right),
\end{equation}
with ${\bf S}_{\alpha,{\bf r}} \cdot {\bf S}_{\alpha,{\bf r}} = \left( S_{\alpha,{\bf r}}^0 \right)^2 - \left( S_{\alpha,{\bf r}}^x \right)^2 - \left( S_{\alpha,{\bf r}}^y \right)^2 -\left( S_{\alpha,{\bf r}}^z \right)^2$, $S_{\alpha,{\bf r}}^{\upsilon} = \dfrac{1}{2} c_{\alpha,{\bf r},\beta}^\dagger \sigma_{\beta,\gamma}^{\upsilon \phantom\dagger} c_{\alpha,{\bf r},\gamma}^{\phantom\dagger} $ and $\sigma_{\beta,\gamma}^{0 \phantom\dagger}$ is the identity and $\sigma_{\beta,\gamma}^{\upsilon \phantom\dagger}, \, \upsilon = \{x,y,z\}$ are the Pauli matrices. Using a mean field approximation, \textit{i.e.} neglecting the terms of order $\left[{\bf S}_{\alpha,{\bf r}} - \langle {\bf S}_{\alpha,{\bf r}} \rangle \right]^2$, we find
\begin{equation}
{\cal H}_U \approx \dfrac{U}{2} \sum_{{\bf r}, \alpha} \left[ 2 {\bf S}_{\alpha,{\bf r}} \cdot \langle {\bf S}_{\alpha,{\bf r}} \rangle - \langle {\bf S}_{\alpha,{\bf r}} \rangle \cdot \langle {\bf S}_{\alpha,{\bf r}} \rangle + S_{\alpha,{\bf r}}^{0} \right].
\end{equation}
We define $\langle {\bf S}_{\alpha,{\bf r}} \rangle = - {\boldsymbol{\phi}}_{\alpha,{\bf r}}$. We have
\begin{equation}
{\cal H}_U \approx - U \sum_{{\bf r}, \alpha}  \left( {\bf S}_{\alpha,{\bf r}} \cdot {\boldsymbol{\phi}}_{\alpha,{\bf r}} + \dfrac{1}{2} {\boldsymbol{\phi}}_{\alpha,{\bf r}} \cdot {\boldsymbol{\phi}}_{\alpha,{\bf r}} \right) +\dfrac{U}{2} \sum_{{\bf r}, \alpha}  S_{\alpha,{\bf r}}^{0}.
\end{equation}
We assume that the field ${\boldsymbol{\phi}}_{\alpha,{\bf r}}$ has the translation symmetry of the lattice, \textit{i.e.} ${\boldsymbol{\phi}}_{\alpha,{\bf r}} = {\boldsymbol{\phi}}_{\alpha,{\bf r} + {\bf e}_\kappa}$, with $\kappa=1,2$.
Based on that we have ${\boldsymbol{\phi}}_{\alpha,{\bf r}} = {\boldsymbol{\phi}}_{\alpha}$. Using Fourier transformation, we get
\begin{equation}
{\cal H}_U \approx - U \sum_{{\bf k}, \alpha} \left( {\boldsymbol{\phi}}_{\alpha} \cdot {\bf S}_{\alpha,{\bf k}} - \dfrac{S_{\alpha,{\bf k}}^{0}}{2} \right)  -  \dfrac{U N}{2} \sum_{\alpha} {\boldsymbol{\phi}}_{\alpha} \cdot {\boldsymbol{\phi}}_{\alpha} ,
\end{equation}
with $\bf k$ the momentum space variable and $N=N_1 N_2$ the total number of unit cells. We write 
\begin{equation}
{\cal H}_U \approx  \sum_{{\bf k}} \psi_{\bf k}^\dagger {\cal H}_{\textrm{int}} \psi_{\bf k}^{\phantom \dagger} -  \dfrac{U N}{2} \sum_{\alpha} {\boldsymbol{\phi}}_{\alpha} \cdot {\boldsymbol{\phi}}_{\alpha},
\end{equation}
with
\begin{subequations}
\begin{align}
{\cal H}_{\textrm{int}} &= \dfrac{U}{4} \mathbb{1} + 
\begin{pmatrix}
H_{\textrm{int}}^{\uparrow,\uparrow} &H_{\textrm{int}}^{\downarrow,\uparrow}\\
H_{\textrm{int}}^{\uparrow,\downarrow} &H_{\textrm{int}}^{\downarrow,\downarrow}
\end{pmatrix} \,, 
\\
H_{\textrm{int}}^{\sigma,\sigma}&=-\dfrac{U}{2} 
\begin{pmatrix}
\phi_{R,\sigma} & 0 & 0\\
0 &  \phi_{B,\sigma}& 0 \\
0 & 0 &  \phi_{G,\sigma} 
\end{pmatrix}\,,
\\
H_{\textrm{int}}^{\downarrow,\uparrow} &=
\left[{H_{\textrm{int}}^{\uparrow,\downarrow}}\right]^\dagger=\dfrac{U}{2}
\begin{pmatrix}
  \phi_{R,-}& 0 &0\\
 0 &\phi_{B,-}& 0 \\
0 & 0 & \phi_G^-
\end{pmatrix}\,.
\end{align}
\end{subequations}
Here 
\begin{subequations}
\begin{align}
\phi_{\alpha,\sigma}&=\phi_\alpha^0
-s_z \phi_\alpha^z\\
\phi_{\alpha,\pm}&=\phi_\alpha^x \pm i \phi_\alpha^y
\end{align} 
\end{subequations}
with $s_z=1~(-1)$ for $\sigma=\uparrow~(\downarrow)$ and 
\begin{equation}
\psi_{\boldsymbol{k}}^\dagger = \left(c_{R,\boldsymbol{k},\uparrow}^\dagger, c_{B,\boldsymbol{k},\uparrow}^\dagger, c_{G,\boldsymbol{k},\uparrow}^\dagger,c_{R,\boldsymbol{k},\downarrow}^\dagger, c_{B,\boldsymbol{k},\downarrow}^\dagger, c_{G,\boldsymbol{k},\downarrow}^\dagger \right).
\end{equation}
We also write ${\cal H}_{0}$ in momentum space. 
\begin{subequations}
\begin{eqnarray}
\label{Hk_matrix}
&&\hspace{-0.75cm}{\cal H}_{0}=\sum_{{\bf k}}\psi_{{\bf k}}^{\dagger}\left(
\begin{array}{cc}
{\cal H}_{\uparrow}({\bf k}) & 0 \\
0 & {\cal H}_{\downarrow}({\bf k})\\
\end{array}
\right)\psi_{{\bf k}}^{\phantom\dagger}, \\
&&\hspace{-0.75cm}{\cal H}_{\sigma}({\bf k})\hspace{-0.1cm}=\hspace{-0.1cm}\left(
\hspace{-0.1cm}
\begin{array}{ccc}
\lambda & \varepsilon_1({\bf k}) & \varepsilon_2({\bf k}) \\
\varepsilon_1({\bf k}) & 0 & e^{i s_z\phi}\varepsilon_3({\bf k}) \\
\varepsilon_2({\bf k}) &e^{-i s_z\phi}\varepsilon_3({\bf k})& 0\\
\end{array}
\hspace{-0.1cm}
\right) ,
\end{eqnarray}
\end{subequations}
with $\varepsilon_\gamma({\bf k}) = -2t \cos (\frac{1}{2}{\bf k} \cdot {\bf e}_\gamma)$, where $\gamma=1,2,3$ and ${\bf e}_3={\bf e}_2-{\bf e}_1$.
We note that we develop this method for $\gamma=0$. 
The reason is that since the basis of the Hilbert space is written in $\sigma=\uparrow$ and $\sigma=\downarrow$ fermionic states, the Hamiltonian for $\gamma=0$ is block diagonal and easier to handle.  For $\gamma \neq 0$ the Hamiltonian is no longer block diagonal and rotation of the fermionic operators is necessary to make the Hamiltonian block diagonal. Accordingly, the calculations are much more complicated.

Taking both terms, ${\cal H}_U$ and ${\cal H}_{0}$, into account we obtain
\begin{equation}
{\cal H} \approx \sum_{{\bf k}} \psi_{\bf k}^\dagger h({\bf k}) \psi_{\bf k}^{\phantom \dagger} -  \dfrac{U N}{2} \sum_{\alpha=R,B,G} {\boldsymbol{\phi}}_{\alpha} \cdot {\boldsymbol{\phi}}_{\alpha},
\end{equation}
with 
\begin{subequations}
\begin{align}
&h({\bf k}) = \dfrac{U}{4} \mathbb{1} + 
\begin{pmatrix}
h^{\uparrow,\uparrow} &h^{\downarrow,\uparrow}\\
h^{\uparrow,\downarrow} &h^{\downarrow,\downarrow}
\end{pmatrix} \, ,
\\
&h^{\sigma,\sigma} =  \begin{pmatrix}
\lambda + \dfrac{U}{2}\phi_{R,\sigma} & \varepsilon_1({\bf k}) & \varepsilon_2({\bf k})\\
\varepsilon_1({\bf k}) & \dfrac{U}{2}\phi_{B,\sigma} & e^{i\phi}\varepsilon_3({\bf k})  \\
\varepsilon_2({\bf k}) & e^{-i\phi}\varepsilon_3({\bf k}) & \dfrac{U}{2}\phi_{G,\sigma} 
\end{pmatrix} \,,
\\
&h^{\downarrow,\uparrow} =\left[h^{\uparrow,\downarrow}\right]^\dagger=
\begin{pmatrix}
  \dfrac{U}{2}\phi_{R,-} & 0 &0\\
0 &\dfrac{U}{2}\phi_{B,-} & 0 \\
 0 & 0 & \dfrac{U}{2} \phi_{G,-} 
\end{pmatrix} \,. 
\nonumber
\end{align}
\end{subequations}

Our goal is to determine the set of parameters $\phi_\alpha^\upsilon$ ($\upsilon=x,y,z$) which minimize the ground state energy $E_{\bf k}$ associated to the Hamiltonian ${\cal H}$. We are interested in the limit $\lambda \gg t$ and $\lambda \gg U$ for which the computation is tractable.

\section{Results without staggered potential}
\label{Results_without}

As mentioned above, in this work we consider two different setups. 
First we consider the case where the onsite energies are independent of $\bf r$ and are non-zero only on $R$ sublattice sites, i.e.  $V_{{\bf r},R}=\lambda$ and $V_{{\bf r},B}=V_{{\bf r},G}=0$. We perform calculations for $\gamma=0$ and for $n=2/3$ filling.  We mainly perform calculations for $\phi=\pi/2$ using R-DMFT and we investigate the behavior of the system for $\lambda \gg t$ using both analytical methods method introduced in Secs. \ref{Perturbation_lambdaR_U} and \ref{Stochastic_MF}.

\begin{figure}[t!]
\centering{\includegraphics[width=0.95\columnwidth]{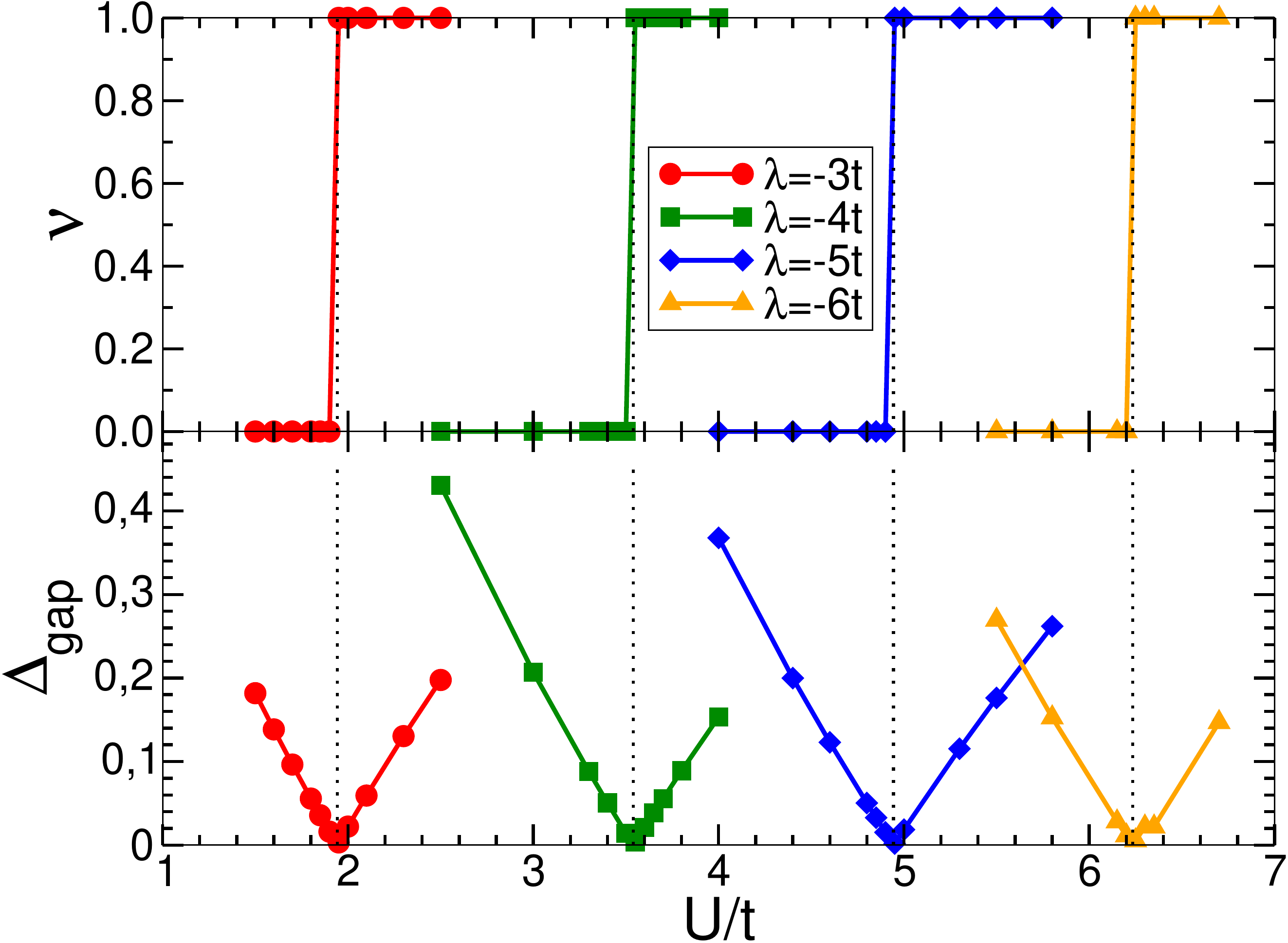}}
\caption{The $\mathbb{Z}_2$ number $\nu$ (upper panel) and gap $\Delta_{\rm gap}$ (lower panel) as a function of the local Hubbard interaction $U$ for different values of the onsite energies. Vertical dotted lines correspond to the phase transition between band and topological insulator. 
Other parameters are $n=2/3$, $V_{{\bf r},R}=\lambda$ and $V_{{\bf r},B}=V_{{\bf r},G}$, $\gamma=0$, and $\phi=\pi/2$. 
}
\label{Fig:TI-BI_transition_R}
\end{figure}

\subsection{$\phi=\pi/2$}
\label{DMFT_calc_2d3}

In our previous work, in Ref.~\onlinecite{ti.le.21}, we studied the current model without Hubbard interaction. Among other parameters, we studied a $2/3$-filled system for $\phi=\pi/2$, where the onsite energies are nonzero only in $R$-sublattice sites, and we constructed a $\gamma-\lambda$ phase diagram. We obtained three distinguishable phases: (i)~band insulator for $\lambda<-2t$, (ii)~topological insulator for $\lambda>-2t$ and $\gamma \lessapprox 0.1$, and (iii)~metallic phase for $\lambda>-2t$ and $\gamma \gtrapprox 0.1$. Here we perform calculations for temperature $T=0.1t$ and $\gamma=0$. We investigate the effect of the Hubbard interaction $U$ using R-DMFT. To obtain the desired filling $n=2/3$, we should adjust the chemical potential $\mu$. In the calculations presented below, we consider the system size $N_1=N_2=20$. 

\begin{figure}[t]
\centering{\includegraphics[width=0.95\columnwidth]{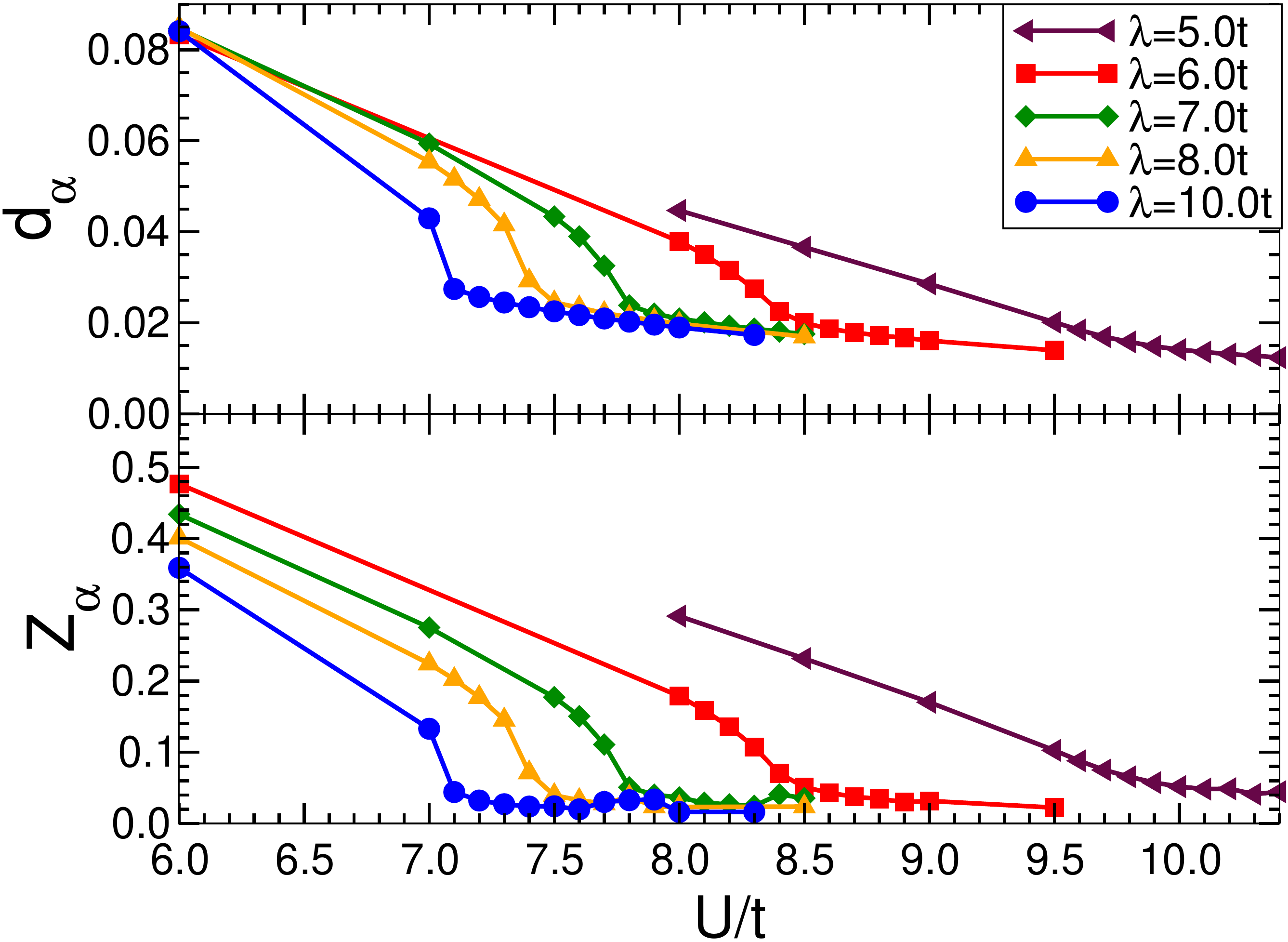}}
\caption{The double occupancy $d_B \simeq d_G$ (upper panel) and quasi-particle weight $Z_B \simeq Z_G$ (lower panel) as a function of local Hubbard interaction $U$ for different values of onsite energies and temperature $T=0.1t$. 
Other parameters are the same as in Fig. \ref{Fig:TI-BI_transition_R}.
}
\label{Fig:TI-MI_transition_R}
\end{figure}

First, we study the transition from the band insulator to the topological insulator. For this purpose, we calculate the $\mathbb{Z}_2$ number $\nu$ and the gap $\Delta_{\rm gap}$ as a function of the local Hubbard interaction $U$ for different values of the onsite energies $\lambda<-2t$ (see Fig.~\eqref{Fig:TI-BI_transition_R}). For these values of the onsite energies, the system in the non-interacting limit is in the band-insulator phase. Thus, for small values of $U$ the $\mathbb{Z}_2$ number is $\nu=0$ and the gap $\Delta_{\rm gap}>0$. As the interaction strength $U$ increases, the gap $\Delta_{\rm gap}$ decreases and for a certain critical value $U=U_c$ the gap closes and after further increase of $U$ the gap opens again and the $\mathbb{Z}_2$-number is $\nu=1$. So the transition to the topological insulator phase takes place. 
For $\lambda>-2t$, the system is in the topological insulator phase and no topological transition was observed with increasing interaction strength.

As it was already discussed in Ref. \onlinecite{ti.le.21}, for onsite energies applied only to $R$ sublattice sites, in the limiting case $\lambda \gg t$ the filling of $R$ sublattice sites is $n_R \ll 1$ and  the system can be described by an effective half-filled model defined on a square lattice with alternating diagonal hopping.  
Therefore, we expect that in this case for large Hubbard interactions $U \gg t$ the transition to the Mott insulator phase takes place, which we further investigate.  For this purpose we perform calculations in the paramagnetic phase and study the quasi-particle weight $Z_{{\bf r},\alpha}=Z_\alpha$ and the double occupancy $d_{{\bf r},\alpha}=d_\alpha$ as a function of $U$ for large onsite energies.

\begin{figure}[t]
\centering{\includegraphics[width=0.95\columnwidth]{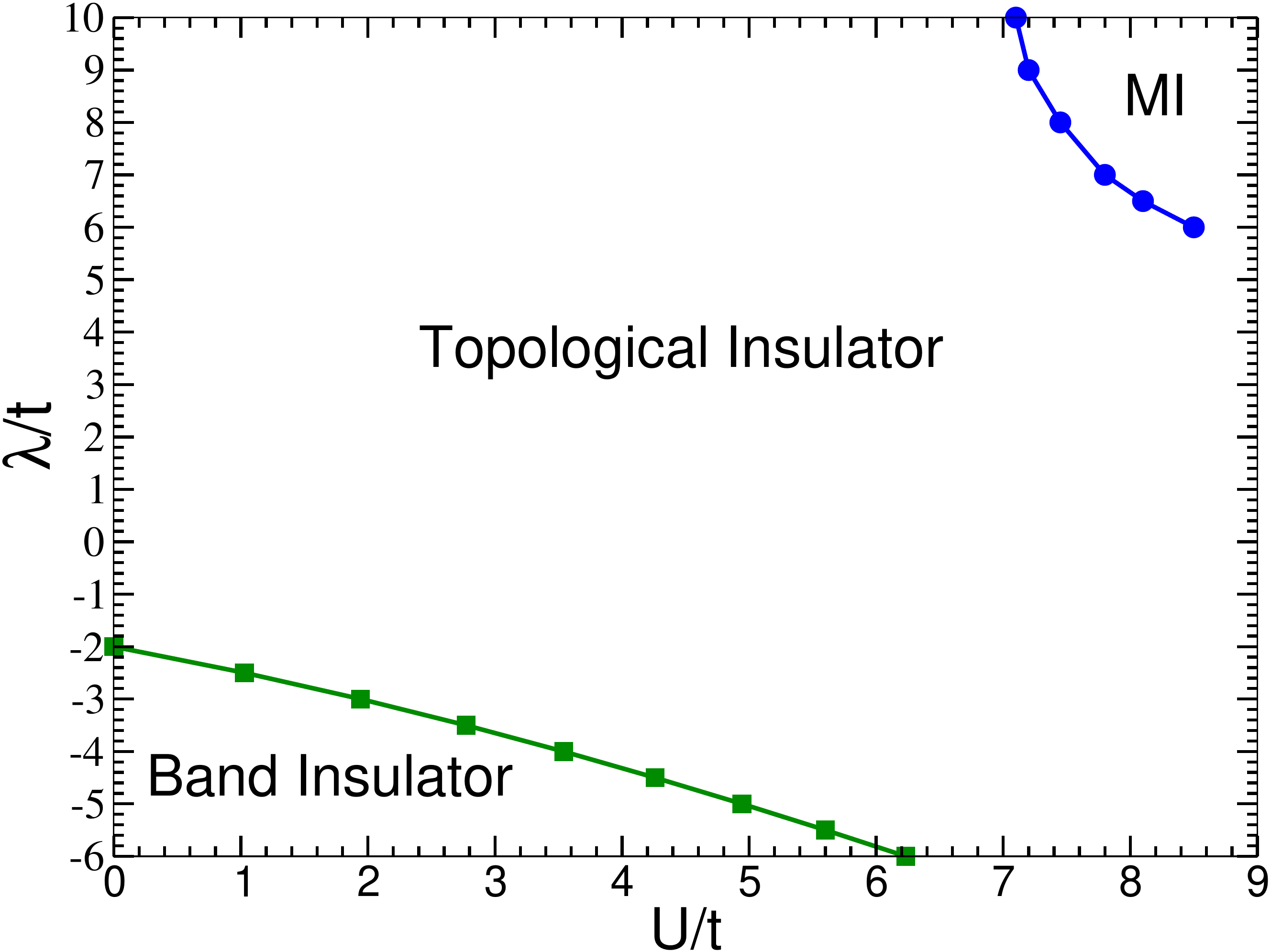}}
\caption{The paramagnetic phase diagram for $n=2/3$ filled system. The green curve with squares corresponds to a transition from band insulator to topological insulator, while the blue curve with circles corresponds to a transition to the Mott insulator phase. 
The temperature is $T=0.1t$ and other parameters are the same as in Fig. \ref{Fig:TI-BI_transition_R}. 
}
\label{Fig:Phased-diagram_R_MI}
\end{figure}
\begin{figure}[t]
\centering{\includegraphics[width=0.95\columnwidth]{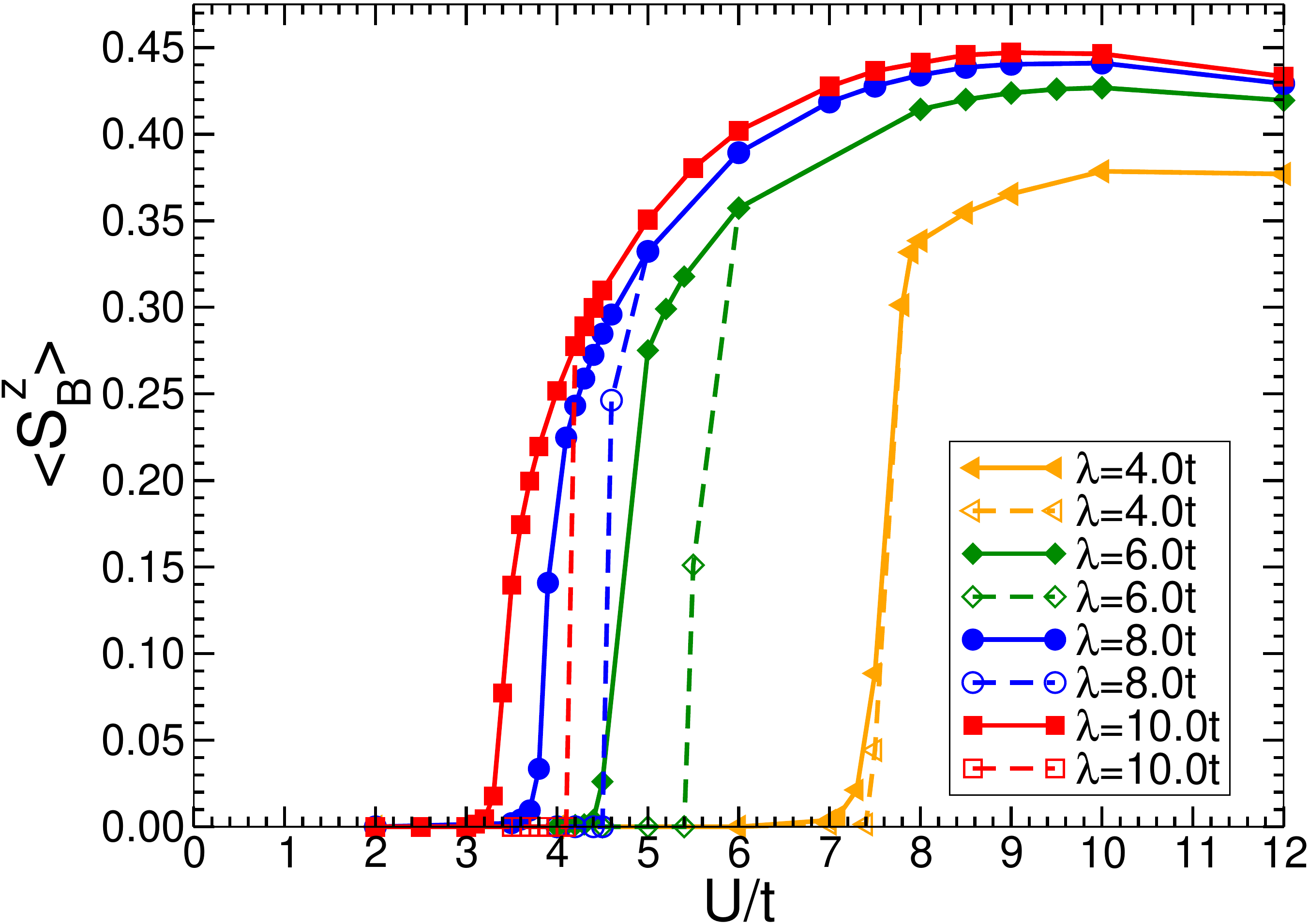}}
\caption{$\langle S_B^z\rangle$ for $B$ sublattice sites ($\langle S_B^z \rangle = -\langle S_G^z\rangle$) as a function of local Hubbard interaction $U$ for different values of the onsite energies. For solid (dashed) lines with closed (open) symbols initial self-energy is magnetic (paramagnetic). 
The temperature is $T=0.1t$ and other parameters are the same as in Fig. \ref{Fig:TI-BI_transition_R}.
}
\label{Fig:AF_transition_R}
\end{figure}
\begin{figure}[t]
\centering{\includegraphics[width=0.95\columnwidth]{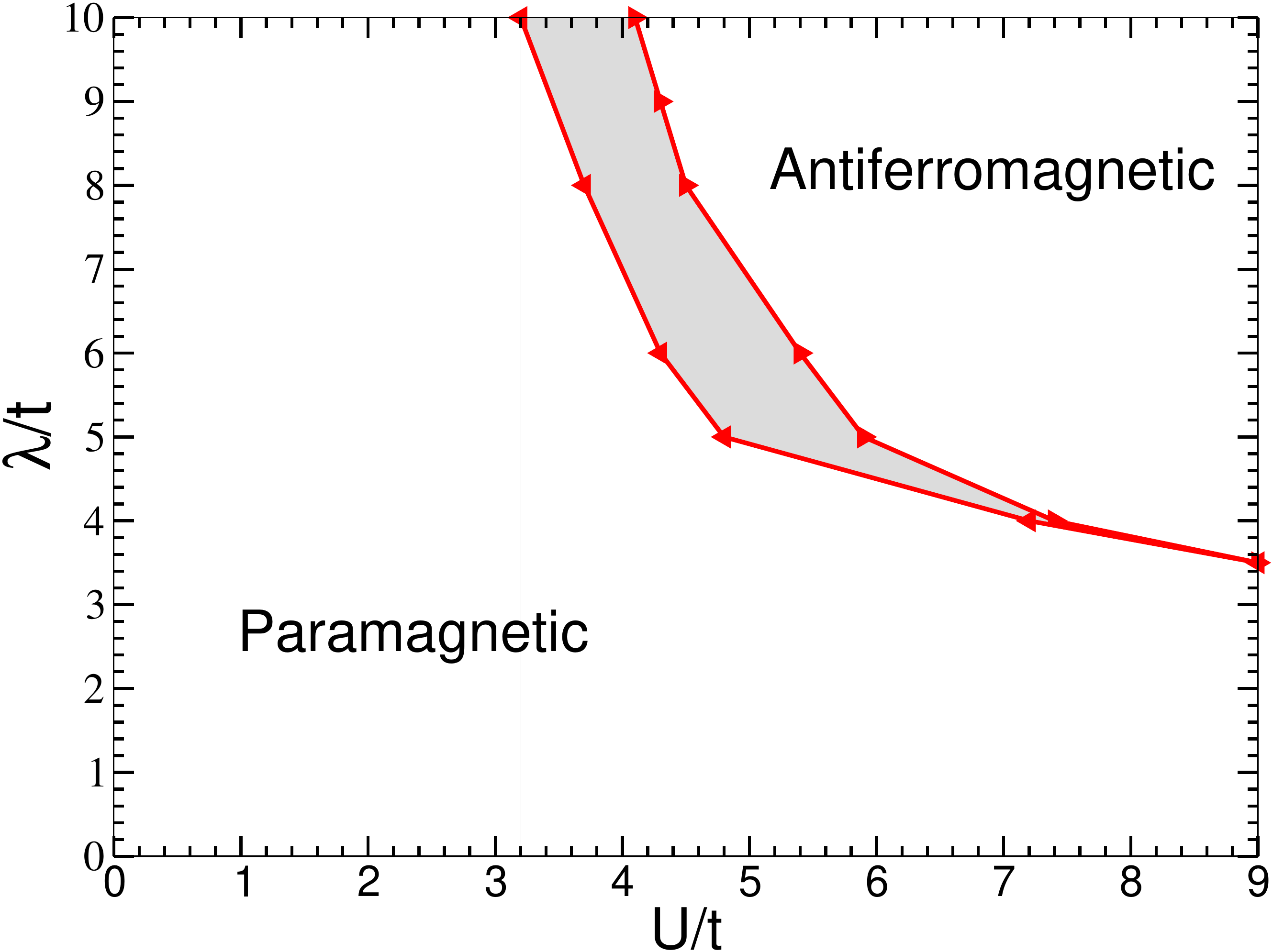}}
\caption{Magnetic phase diagram for filling $n=2/3$. 
The red curves with triangles separate the paramagnetic phase for weak interactions from the antiferromagnetic phase for strong interactions. We obtain a hysteresis region (gray area between red curves). The curve with left (right) triangles is obtained when we start DMFT iterations from an antiferromagnetic (a paramagnetic) initial self-energy.   
The temperature is ${T=0.1t}$ and other parameters are the same as in Fig.~\ref{Fig:TI-BI_transition_R}. 
}
\label{Fig:Phased-diagram_R_AF}
\end{figure}

Our calculation indeed shows that for large onsite energies filling of $R$-sublattice sites is very small, $n_R \ll 1$ and correspondingly $Z_R \simeq 1$, and $d_R \ll 1$ for all interaction strengths. More interesting is the behavior of double occupancy and quasi-particle weight for $B$ and $G$ sublattice sites.  
Our results for different values of the onsite energies are shown in Fig.~\ref{Fig:TI-MI_transition_R}.  
We observe that $d_B \simeq d_G$ and $Z_B \simeq Z_G$ decrease with increase of $U$. 
For $\lambda \gtrapprox 6t$, the phase  transition to the Mott insulator phase can be identified by a cusp in $d_\alpha$ and $Z_\alpha$ as a function of~$U$. For $\lambda \lessapprox 6t$ we no longer observe a cusp and it seems that considered temperature $T=0.1t$ is high enough and a crossover occurs instead of the phase transition.

To check whether there is hysteresis across the Mott transition curve, we perform calculations starting from two different initial conditions, one metallic and other insulating. Our calculations give the same results. So it seems that either there is no hysteresis or its width is below our accuracy.

Our results are summarized in the phase diagram in Fig.~\ref{Fig:Phased-diagram_R_MI}.  Thus, we obtained three distinguishable phases: the band insulator for negative and large onsite energies and weak interaction, the Mott insulator for large positive onsite energies and strong interaction, and the topological insulator in between.

Finally, we also perform calculations where we remove the paramagnetic constraint. To investigate the magnetic properties of the system, we calculate 
${\langle{\bf S}_\alpha\rangle = \langle c_{{\bf r},\alpha,\sigma}^\dagger \boldsymbol{\sigma}_{\sigma\sigma'}
c_{{\bf r},\alpha,\sigma'}^{\phantom\dagger} \rangle}$ 
for different sublattice sites $\alpha=R,G,B$. Here $\boldsymbol{\sigma}_{\sigma\sigma'}=(\sigma_{\sigma\sigma'}^x,\sigma_{\sigma\sigma'}^y,\sigma_{\sigma\sigma'}^z)$ are the Pauli matrices.  
We obtain that ${\langle S_{\alpha=R,B,G}^x\rangle \simeq \langle S_{\alpha}^y\rangle \simeq \langle S_R^z \rangle \simeq 0}$ and the only nontrivial magnetization is along the $z$ direction. We have $\langle S_B^z\rangle=-\langle S_G^z\rangle$.
In Fig.~\ref{Fig:AF_transition_R} we plot $\langle S_B^z \rangle $ as a function of $U$ for different values of onsite energies $\lambda$. We obtain that for weak interactions $\langle S_B^z \rangle =0$ and the system is in the paramagnetic phase. 
Obviously results obtained in this limit are the same as discussed above, when we force system to be paramagnetic. 
Here we also would like to note that as the interaction increases, for $U > U_c(\lambda)$ we obtain a finite value of ${\langle S_B^z \rangle}$ and the transition to the antiferromagnetic phase takes place. 
With further increase of the interaction ${\langle S_B^z \rangle}$ reaches a maximum and after further increase of the interaction ${\langle S_B^z \rangle}$ slowly decreases.

To further investigate the phase transition, we perform calculations with different initial self-energies: a paramagnetic and a magnetic one. For strong onsite energies near the phase transition, we obtain two different solutions, the paramagnetic and the magnetic solution, depending on whether we start from the paramagnetic or the magnetic self-energy. So we  obtain a hysteresis region (gray area in Fig. \ref{Fig:Phased-diagram_R_AF}). Depending on we start from the magnetic self-energy or the paramagnetic self-energy we obtained two different solutions in this region.

Our results are summarized in Fig.~\ref{Fig:Phased-diagram_R_AF}, where we show the phase transition curve between the paramagnetic and the magnetic phases. For weak interactions the system is in the paramagnetic phase, while for strong interactions the system is in the antiferromagnetic phase. We also obtain hysteresis region (gray area between two red curves).

\subsection{$\lambda \gg t$ and $U \gg t$}
\label{results_for_large_lambda_and_U}

In the limit $\lambda \gg t$ and $U \gg t$ and considering $n=2/3$ filling, the ground states of ${\cal H}_U + {\cal H}_\lambda$ are the states, that we denote $\ket{\psi}$, with no particles on the $R$-sublattice sites and one particle at each $B$ and $G$ sublattice sites. These states are linear combinations of the $\sim 2^{2N}$ states ${\ket{0,\sigma_{1,B},\sigma_{1,G}, \dots, 0,\sigma_{N,B},\sigma_{N,G}}}$, with $\sigma_{i,\alpha}= \{\uparrow,\downarrow\},$ for $i \in [1, \dots, N]$ and $\alpha=B,G$.
The first order terms of the expansion, $\bra{\psi'} {{\cal H}_t}\ket{\psi}$ vanish. Here $\ket{\psi}$ and $ \ket{\psi'}$ belong to the ground state subspace, that we denote $S_0$. Indeed, ${\cal H}_t \ket{\psi}$ is composed of states which are linear combination of states with 2 particles on a color $B$ or $G$ site or 1 particle on a color $R$ site. Theses states are orthogonal to $\ket{\psi'}$. 
The second order terms give the effective Hamiltonian (at order 2), that we denote ${\cal H}_{\textrm{eff}}$
\begin{equation}
\bra{\psi'} {{\cal H}_{\textrm{eff}}}\ket{\psi}= \sum_{\ket{m}\notin S_0} \dfrac{\bra{\psi'} {{\cal H}_t} \ket{m} \bra{m} {{\cal H}_t} \ket{\psi}}{-E_m},
\end{equation}
with $\ket{m}$ the eigenstates of ${\cal H}_U + {\cal H}_\lambda$ which do not belong to the ground state subspace $S_0$ and $E_m$ is the  corresponding eigenenergy. Here we note that the ground state energy is ${E_0=0}$. The numerator in the above equation is composed of terms like $\bra{\psi'} c_{i \sigma}^\dagger c_{j \sigma}\ket{m} \bra{m} c_{k \sigma'}^\dagger c_{l \sigma'} \ket{\psi}$ with $\langle i,j \rangle$, and $\langle k,l \rangle$ two pairs of nearest neighbors in the lattice. These are non vanishing only if $\ket{m} = c_{k \sigma'}^\dagger c_{l \sigma'} \ket{\psi}$ and $\ket{m} = c_{j \sigma}^\dagger c_{i \sigma} \ket{\psi'}$ which means $k=j$ and $l=i$, because $ \ket{\psi}$ and $ \ket{\psi'}$ are linear combinations of states which all are associated to no particle on the $R$ sublattice sites and exactly one particle to each $B$ and $G$ sublattice site. The energy $E_m$ associated with the intermediate states $\ket{m}$ is $\lambda$ if $j$ is the position of a $R$ sublattice site and $U$ if $j$ is the position of a $B$ or $G$ sublattice site. We notice that if $i$ is associated to a $R$ sublattice site, then the term $\bra{\psi'} c_{i \sigma}^\dagger c_{j \sigma}\ket{m} \bra{m} c_{k \sigma'}^\dagger c_{l \sigma'} \ket{\psi}$ is vanishing.

Further we use the fermionic anti-commutation relations and the fact that $\langle \hat{n}_{i \uparrow} + \hat{n}_{i \downarrow} \rangle = 1$ for $i \in B, G$  and $\langle \hat{n}_{j \uparrow}\rangle =  \langle  \hat{n}_{j \downarrow} \rangle = 0$, for $j \in R$. Here $\langle \mathcal{O} \rangle$ is the eigenvalue of the operator $\mathcal{O}$ when acting on the ground state. 
Perturbation due to the processes including $R$ sublattice site gives terms proportional to $t^2/\lambda$ which are constant energy terms and we obtain 
\begin{align} \label{bbb}
{\cal H}_{\textrm{eff}} = &\dfrac{4 t^2}{U}\sum_{{\bf r}} \bigg[ S_{B, {\bf r}}^zS_{G, {\bf r}}^z + S_{B, {\bf r}+{\bf e}_3}^zS_{G, {\bf r}}^z \nonumber \\ &+ \cos 2 \phi \bigg( S_{B, {\bf r}}^x S_{G, {\bf r}}^x +S_{B, {\bf r}}^y S_{G, {\bf r}}^y \nonumber \\ &+ S_{B, {\bf r}+{\bf e}_3}^x S_{G, {\bf r}}^x+ S_{B, {\bf r}+{\bf e}_3}^y S_{G, {\bf r}}^y\bigg) \nonumber \\ &+ \sin 2 \phi \bigg( S_{B, {\bf r}}^x S_{G, {\bf r}}^y -S_{B, {\bf r}}^y S_{G, {\bf r}}^x \nonumber \\ &+ S_{B, {\bf r}+{\bf e}_3}^x S_{G, {\bf r}}^y- S_{B, {\bf r}+{\bf e}_3}^y S_{G, {\bf r}}^x\bigg)\bigg],
\end{align}
where $S_{\alpha,{\bf r}}^{r} = \dfrac{1}{2} c_{\alpha,{\bf r},\beta}^\dagger \sigma_{\beta,\gamma}^{r \phantom\dagger} c_{\alpha,{\bf r},\gamma}^{\phantom\dagger} $ and $\sigma_{\beta,\gamma}^{r \phantom\dagger}, \, r = \{x,y,z\}$ are the Pauli matrices.

First, when $\phi =0$, the effective Hamiltonian is the one of decoupled antiferromagnetic 1D Heisenberg spin chains. This has been studied a lot, \textit{e.g.} using the bosonization technique, and it is known to be characterized by an algebraic decay of the spin correlation function\cite{giam.03, go.ne.04}. 

When $\phi =\pi/2$, the transformation (that preserves the commutation relations)
\begin{equation}
S_{B,{\bf r}}^x \rightarrow - S_{B,{\bf r}}^x,\,\, S_{B,{\bf r}}^y \rightarrow - S_{B,{\bf r}}^y,\,\, \textrm{and} \,\, S_{B,{\bf r}}^z\rightarrow S_{B,{\bf r}}^z, 
\end{equation}
on the spin operators at the $B$-sublattice site in each $G$-$B$ chain gives back the effective Hamiltonian at $\phi =0$. We deduce that the $z$ antiferromagnetic and $xy$ ferrromagnetic effective Hamiltonian at $\phi =\pi/2$ is also characterized by an algebraic decay of the spin correlation function.

At $\phi \neq \{0, \pi/2 \}$ the effective Hamiltonian possesses XXZ anisotropy and also contains Dzyaloshinskii-Moriya interaction, $S_{i}^x S_{j}^y-S_{i}^y S_{j}^x$, where $i$ and $j$ are the $B$ and $G$ nearest neighbors. Here we want to compute the ground state in the classical limit of large spin $S$  We write the value of the spin operators in the classical ground state $\langle S_{\alpha,{\bf r}}^{x} \rangle = \rho_{\alpha,{\bf r}} \sin \theta_{\alpha,{\bf r}} \cos \varphi_{\alpha,{\bf r}} $, $\langle S_{\alpha,{\bf r}}^{y} \rangle = \rho_{\alpha,{\bf r}} \sin \theta_{\alpha,{\bf r}} \sin \varphi_{\alpha,{\bf r}} $, and $\langle S_{\alpha,{\bf r}}^{z} \rangle = \rho_{\alpha,{\bf r}} \cos \theta_{\alpha,{\bf r}} $, with $\theta_{\alpha,{\bf r}} \in [0,\pi]$, and $\varphi_{\alpha,{\bf r}} \in [0,2\pi]$ and $\rho_{\alpha,{\bf r}}$ is the norm of the spins in the classical ground state. 
In this limit, we obtain
\begin{align} 
{\cal H}_{\textrm{eff}} = &\dfrac{4 t^2}{U}\sum_{{\bf r}} \rho_{B,{\bf r}} \rho_{G,{\bf r}}\big[  \cos \theta_{B,{\bf r}}  \cos \theta_{G,{\bf r}}  \\ 
&+ \sin \theta_{B,{\bf r}}  \sin \theta_{G,{\bf r}} \cos \left[ 2 \phi - (\varphi_{G,{\bf r}} - \varphi_{B,{\bf r}}) \right]  \big]\nonumber  \\ &+\rho_{B,{\bf r}+{\bf e}_3} \rho_{G,{\bf r}}\big[ \cos \theta_{B, {\bf r}+{\bf e}_3}  \cos \theta_{G,{\bf r}} \nonumber \\ &+ \sin \theta_{B, {\bf r}+{\bf e}_3}  \sin \theta_{G,{\bf r}} \cos \left[ 2 \phi - (\varphi_{G,{\bf r}} - \varphi_{B, {\bf r}+{\bf e}_3})\right]  \big],
\nonumber 
\end{align}
We obtain a similar effective Hamiltonian for the Hofstadter-Hubbard model on a square lattice\cite{or.co.13}. 

The classical ground state minimizing the energy associated to ${\cal H}_{\textrm{eff}} $ is associated to $\rho_{B,{\bf r}} =\rho_{B,{\bf r}+{\bf e}_3}=\rho_{G,{\bf r}}$, $\theta_{B,{\bf r}} = \theta_{B, {\bf r}+{\bf e}_3} = \pi -  \theta_{G,{\bf r}}$ and $\varphi_{G,{\bf r}} - \varphi_{B,{\bf r}} =  \varphi_{G,{\bf r}} - \varphi_{B, {\bf r}+{\bf e}_3} = 2 \phi - \pi$. This is in agreement with the results mentioned above. Indeed, in the limit $\phi =0$ and $\phi =\pi/2$, this is the classical ground state of respectively the antiferromagnetic Heisenberg spin chain and the $z$ antiferromagnetic and $xy$ ferromagnetic spin chain.

\subsection{$\lambda \gg t$ and $\lambda \gg U$}
\label{analytical_calc}

We write the wave function $\psi_{\bf k}$ and correspondingly the Hamiltonian $h({\bf k})$ as follows
${\tilde\psi_{\boldsymbol{k}}^\dagger = \left(c_{R,\boldsymbol{k},\uparrow}^\dagger, c_{R,\boldsymbol{k},\downarrow}^\dagger, c_{B,\boldsymbol{k},\uparrow}^\dagger,c_{B,\boldsymbol{k},\downarrow}^\dagger, c_{G,\boldsymbol{k},\uparrow}^\dagger, c_{G,\boldsymbol{k},\downarrow}^\dagger \right)}$ and 
\begin{equation}
\tilde{h}({\bf k}) =\dfrac{U}{4}\mathbb{1} +  
\begin{pmatrix}
 \tilde{h}_R({\bf k}) &\tilde{h}_\Delta({\bf k}) \\
\tilde{h}^\dagger_\Delta({\bf k})  & \tilde{h}_0({\bf k})
\end{pmatrix} \,.
\end{equation}
Here 
\begin{subequations}
\begin{align}
&\tilde{h}_R({\bf k}) =
\begin{pmatrix}
\lambda - \dfrac{U}{2}\phi_{R,\uparrow} & \dfrac{U}{2} \phi_{R,-} \\
\dfrac{U}{2} \phi_{R,+} & \lambda - \dfrac{U}{2} \phi_{R,\downarrow} \\
\end{pmatrix},
\\
&\tilde{h}_\Delta({\bf k})  =
\begin{pmatrix}
\varepsilon_1({\bf k})&  0& \varepsilon_2({\bf k}) &0\\
0 & \varepsilon_1({\bf k}) &0 & \varepsilon_2({\bf k}) \\
\end{pmatrix},
\\
&\tilde{h}_0({\bf k}) = 
\begin{pmatrix}
-\dfrac{U}{2}\phi_{B,\uparrow} & \dfrac{U}{2} \phi_{B,-} & 
e^{i\phi}\varepsilon_3({\bf k}) & 0 \\
\dfrac{U}{2} \phi_{B,-} &
- \dfrac{U}{2} \phi_{B,\downarrow} &
0 & e^{-i\phi}\varepsilon_3({\bf k}) \\
e^{-i\phi}\varepsilon_3({\bf k}) & 0 & 
-\dfrac{U}{2} \phi_{G,\uparrow} & \dfrac{U}{2} \phi_{G,-} \\
0& e^{i\phi}\varepsilon_3({\bf k}) & \dfrac{U}{2} \phi_{G,+} &
- \dfrac{U}{2} \phi_{G,\downarrow} 
\end{pmatrix}.
\end{align}
\end{subequations}

We want to determine the four lowest eigenvalues $X$ associated to $\tilde{h({\bf k})}$. The two other eigenvalues are of order $\lambda$ and do not interest us here. We have
\begin{align}
&\textrm{det}\left(\tilde{h}({\bf k}) -\dfrac{U}{4} \mathbb{1} - X \mathbb{1} \right) = \textrm{det}\left( \tilde{h}_R({\bf k}) - X \mathbb{1}\right) \nonumber \\ &\times \textrm{det}\left(\tilde{h}_0({\bf k}) - X \mathbb{1} - \tilde{h}^\dagger_\Delta({\bf k}) \left( \tilde{h}_R({\bf k}) - X \mathbb{1} \right) ^{-1} \tilde{h}_\Delta({\bf k})  \right)
\end{align}
Here $\textrm{det}\left( \tilde{h}_R({\bf k}) - X \mathbb{1}\right) \neq 0$ and we write 
\begin{equation} 
\tilde{h}_{\textrm{eff}}({\bf k}) = \tilde{h}_0({\bf k}) - \tilde{h}^\dagger_\Delta({\bf k}) \left( \tilde{h}_R({\bf k}) - X \mathbb{1} \right) ^{-1} \tilde{h}_\Delta({\bf k}) 
\,. 
\end{equation}
It reads (with implicit ${\bf k}$ dependency)
\begin{align}
&\tilde{h}_{\textrm{eff}} = \tilde{h}_0 \nonumber \\ &- \sum_{\substack{{\sigma,\sigma'} \\ {L,L'}}} \sum_{i=1}^2 \dfrac{\ket{L,\sigma} \bra{L,\sigma}\tilde{h}^\dagger_\Delta \ket{R_{i}} \bra{R_{i}} \tilde{h}_\Delta \ket{L',\sigma'} \bra{L',\sigma'}}{E_{i} - X} 
\end{align}
where the sum runs over the two color B and G for $L$ and $L'$ and $\sigma$ and $\sigma'$ run over both values of the spin degree of freedom. $\ket{R_{i}}$ and $E_{i}$ are respectively the eigenvectors and the eigenvalues of $\tilde{h}_R({\bf k}) $. \\

\begin{figure}
\includegraphics[scale=0.5]{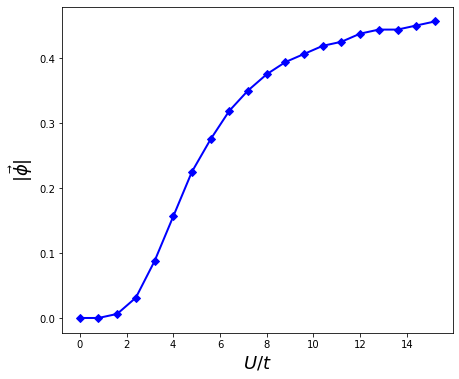}
\caption{Order parameter $| \boldsymbol{ \phi }  |$ computed from the Eq.~\eqref{solu} as a function of $U/t$.}
\label{magorder}
\end{figure}

\subsubsection{The lowest order in $1/\lambda $}
The lowest order in $1/\lambda $ gives
\begin{equation}
\tilde{h}_{\textrm{eff}}({\bf k}) = \tilde{h}_0({\bf k}).
\end{equation}
We assume that $\phi_B^0=\phi_G^0$ because there is no asymmetry in the system justifying that it should be different. We write $\phi^0=\phi_B^0=\phi_G^0$. The equation ${\textrm{det}\left(\tilde{h}_0({\bf k}) - X \mathbb{1} \right) = 0}$ is equivalent to $\textrm{det}\left(\tilde{h}_0({\bf k}) + \dfrac{U}{2} \phi^0 \mathbb{1}  - \tilde{X} \mathbb{1} \right) = 0$ with ${\tilde{X}=X+\dfrac{U}{2} \phi^0}$. It gives $\tilde{X}^4 -h_1({\bf k}) \tilde{X}^2 + h_2({\bf k}) =0 $
with 
\begin{align}
h_1({\bf k}) =\left( \dfrac{U}{2} \right)^2 \left(\boldsymbol{\phi}_B \cdot \boldsymbol{\phi}_B + \boldsymbol{\phi}_G \cdot \boldsymbol{\phi}_G \right) + 2 \varepsilon_3^2({\bf k}),
\end{align}
and
\begin{align}
&h_2({\bf k}) = \left( \dfrac{U}{2} \right)^4 |\boldsymbol{\phi}_B|^2 |\boldsymbol{\phi}_G|^2 + \varepsilon_3^4({\bf k}) \nonumber \\  &- 2 \varepsilon_3^2({\bf k}) \left(\dfrac{U}{2} \right)^2 \phi_B^z \phi_G^z - 2 \varepsilon_3^2({\bf k}) \left(\dfrac{U}{2} \right)^2 \nonumber \\  &\times  \left[ \cos 2 \phi \left(\phi_B^x \phi_G^x+\phi_B^y \phi_G^y \right) + \sin 2 \phi \left(\phi_B^x \phi_G^y-\phi_B^y \phi_G^x  \right) \right].
\end{align}
The solution reads
\begin{align}
X({\bf k}) &= -\dfrac{U}{2} \phi^0 \pm \dfrac{1}{\sqrt{2}} \Bigg\{ 2 \varepsilon_3({\bf k})^2 + \left( \dfrac{U}{2} \right)^2 \left(|\boldsymbol{\phi}_B|^2 + |\boldsymbol{\phi}_G|^2 \right) \nonumber \\ &\pm \dfrac{U}{2} \sqrt{\left( \dfrac{U}{2} \right)^2 \left(|\boldsymbol{\phi}_B|^2 - |\boldsymbol{\phi}_G|^2 \right)^2 + 4 \varepsilon_3({\bf k})^2 g_{\boldsymbol{\phi}}} \, \Bigg\}^{1/2}
\end{align}
with 
\begin{align}
g_{\boldsymbol{\phi}}&= \left( \boldsymbol{\phi}_B + \boldsymbol{\phi}_G \right)^2 - 4 \sin^2 \phi \left(\phi_B^x \phi_G^x + \phi_B^y \phi_G^y \right)
\nonumber\\
&+2 \sin 2 \phi \left(\phi_B^x \phi_G^y-\phi_B^y \phi_G^x\right)
\,. 
\end{align} 
The spectrum associated to $h({\bf k})$ reads $E({\bf k}) = \dfrac{U}{4} + X({\bf k})$. To find the classical magnetic order associated to the system we minimize the total energy
\begin{equation} \label{etot}
E_{\textrm{tot}} =  \sum_i \sum_{{\bf k}} E_i({\bf k}) -  \dfrac{U N}{2} \sum_{\alpha=R,B,G} {\boldsymbol{\phi}}_{\alpha} \cdot {\boldsymbol{\phi}}_{\alpha},
\end{equation}
where the sum over $i$ runs over all the filled bands, which is fixed by the filling factor $n$. In our case  $n=2/3$. It means that the summation in Eq.~\eqref{etot} runs over $i = \{1,2\}$. We look for the set of parameters $\{\phi_B^x, \phi_B^y, \phi_B^z, \phi_G^x, \phi_G^y, \phi_G^z \}$ which minimize $E_{\textrm{tot}} $, by computing the solution of
\begin{equation} \label{eqsol}
\dfrac{\partial E_{\textrm{tot}}}{\partial \phi_L^\upsilon} =\sum_{{\bf k}} \left[ \dfrac{\partial E_1}{\partial \phi_L^\upsilon}+\dfrac{\partial E_2}{\partial \phi_L^\upsilon} \right] + UN \phi_L^\upsilon = 0,
\end{equation}
$L = \{B,G\}$ and $\upsilon = \{x,y,z\}$, associated to the lowest value of $E_{\textrm{tot}} $. This namely yields the following condition
\begin{equation} \label{cond1}
\left( \dfrac{U}{2} \right)^2 \left(|\boldsymbol{\phi}_B|^2 - |\boldsymbol{\phi}_G|^2 \right)^2 + 4 \varepsilon_3({\bf k})^2 g_{\boldsymbol{\phi}}
= 0,
\end{equation}
which does not depend on the value of $U$. Let us write the magnetic order parameters $\phi_\alpha^x = |\boldsymbol{\phi}_{\alpha}| \sin \theta_{\alpha} \cos \varphi_{\alpha} $, $\phi_\alpha^y = |\boldsymbol{\phi}_{\alpha}| \sin \theta_{\alpha} \sin \varphi_{\alpha} $, and $\phi_\alpha^z = |\boldsymbol{\phi}_{\alpha}| \cos \theta_{\alpha} $, with ${|\boldsymbol{\phi}_{\alpha}|  \in [0,1/2]}$, $\theta_{\alpha} \in [0,\pi]$, and $\varphi_{\alpha} \in [0,2\pi]$. In these notations, we have 
\begin{align}
g_{\boldsymbol{\phi}}= &|\boldsymbol{\phi}_B|^2 + |\boldsymbol{\phi}_G|^2+ 2 |\boldsymbol{\phi}_B| |\boldsymbol{\phi}_G| \big( \cos \theta_{B}  \cos \theta_{G}  \nonumber \\ &+ \sin \theta_{B}  \sin \theta_{G} \cos \left[ 2 \phi - (\varphi_{G} - \varphi_{B}) \right]  \big)
\end{align}
Eq. \ref{cond1} gives the condition $|\boldsymbol{\phi}_B| =|\boldsymbol{\phi}_G|$, $\theta_{B} = \pi -  \theta_{G}$ and $\varphi_{G} - \varphi_{B} =  2 \phi - \pi$, which is in agreement with the results we obtained from perturbation theory in the limit $U \gg t$. Besides, here, the minimization of $E_{\textrm{tot}} $ also imposes the following condition on the value of  $| \boldsymbol{ \phi }  |=| \boldsymbol{ \phi }_B  |=| \boldsymbol{ \phi }_G  |$ as a function of $U$
\begin{equation} \label{solu}
\dfrac{1}{U} = \dfrac{1}{2N}\sum_{{\bf k}} \dfrac{1}{\sqrt{4\varepsilon_3^2({\bf k}) + U^2 | \boldsymbol{ \phi }  |^2 } }.
\end{equation}
This condition can be satisfied for all values of $U$ with the appropriate choice of $| \boldsymbol{ \phi }  |$ shown in figure~\ref{magorder}. 

In Sec.~\ref{DMFT_calc_2d3}, using R-DMFT computations in the case $\phi = \pi/2$ and for large enough on-site potential $\lambda$ and Hubbard interaction $U$, we observed the appearance of antiferromagnetic correlations in the $z$ direction. This magnetic order is one of the solutions found from the analytical approach used in this section. It seems that the most general solution (at $\phi = \pi/2$) found from the analytical approach is antiferromagnetic correlations in the $z$ direction and/or ferromagnetic correlations in the $xy$ plane. Nevertheless, it seems that the behavior of the order parameter $| \boldsymbol{ \phi }  |$ found using both approaches is qualitatively the same, but there it shows a small quantitative difference. At the transition, the variation of the magnetization obtained from the R-DMFT method is bigger than the one obtained from the analytical stochastic method.

\begin{figure} 
\includegraphics[scale=0.5]{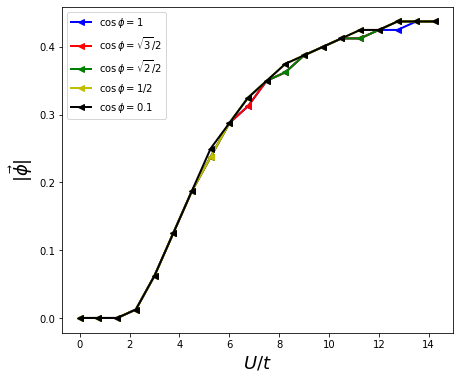}
\caption{Solution $ | \boldsymbol{ \phi }  |$ of Eq.~\eqref{solantif}, as a function of $U/t$, for different values of $\phi$. Here, $\lambda=30t$.}
\label{mago}
\end{figure}

\subsubsection{First order in $1/\lambda $}

Here we investigate the behavior of the order parameter, at order $1/\lambda $, assuming that the solution found at order 0 in $1/\lambda $ ($|\boldsymbol{\phi}_B| =|\boldsymbol{\phi}_G|$, $\theta_{B} = \pi -  \theta_{G}$ and $\varphi_{G} - \varphi_{B} =  2 \phi - \pi$) is still valid. 
Up to the order $t^3/(\lambda \cdot \mbox{max}(U,t))$ and neglecting the terms of order $t^3/\lambda^2$ and $(t^2 U)/\lambda^2$ and higher order terms, we have
\begin{align}
&\textrm{det}\left(\tilde{h}_{\textrm{eff}}({\bf k}) + \dfrac{U}{2} \phi^0 \mathbb{1}  - X \mathbb{1} \right) = \dfrac{h_1/2 - X^2 }{\lambda}  \nonumber \\ &\times \left(-\lambda X^2 -2\left(\varepsilon_1^2 + \varepsilon_2^2 \right)X + \lambda h_1/2 -4 \varepsilon_1  \varepsilon_2 \varepsilon_3 \cos \phi \right)
\end{align}
The solutions to $\textrm{det}\left(\tilde{h}_{\textrm{eff}}({\bf k}) + \dfrac{U}{2} \phi^0 \mathbb{1}  - X \mathbb{1} \right) = 0$ are
\begin{subequations}
\begin{align}
X_1({\bf k}) &= -\sqrt{h_1/2} + \dfrac{2 \varepsilon_1  \varepsilon_2 \varepsilon_3 \cos \phi}{ \lambda \sqrt{h_1/2}} - \dfrac{\left(\varepsilon_1^2 + \varepsilon_2^2 \right)}{ \lambda} ,
\\
X_2({\bf k}) &= - \sqrt{h_1/2} \, ,
\\
X_3({\bf k}) &= \sqrt{h_1/2} - \dfrac{2 \varepsilon_1  \varepsilon_2 \varepsilon_3 \cos \phi}{ \lambda \sqrt{h_1/2}} - \dfrac{\left(\varepsilon_1^2 + \varepsilon_2^2 \right)}{ \lambda} ,
\\
X_4({\bf k}) &=  \sqrt{h_1/2} \, .
\end{align}
\end{subequations}
We have
\begin{equation}
\varepsilon_1^2 + \varepsilon_2^2 \pm \dfrac{2 \varepsilon_1  \varepsilon_2 \varepsilon_3 \cos \phi}{ \sqrt{h_1/2}}>0, \, \forall U >0 \, \forall {\bf k}.
\end{equation}
In the limit $\lambda \gg t$, this leads to
\begin{equation}
X_1({\bf k})<X_2({\bf k})<X_3({\bf k})<X_4({\bf k})
\end{equation}
For each band, the spectrum is $E_i({\bf k}) = \dfrac{U}{4}-\dfrac{U}{2} \phi^0 + X_i({\bf k}), \, i=\{1,2,3,4\}$. The total energy reads
\begin{equation}
E_{\textrm{tot}} = \sum_{i} \sum_{{\bf k}} E_i({\bf k}) -  \dfrac{U N}{2} \sum_{\alpha=R,B,G} {\boldsymbol{\phi}}_{\alpha} \cdot {\boldsymbol{\phi}}_{\alpha},
\end{equation}
where the sum over $i$ runs over all the filled bands, which is fixed by the filling factor $n$. In our case, as we already mentioned $n=2/3$, which means that both lowest energy bands are filled, giving the following total energy
\begin{align}
E_{\textrm{tot}} = &-2\sqrt{h_1/2} + \dfrac{2 \varepsilon_1  \varepsilon_2 \varepsilon_3 \cos \phi}{ \lambda \sqrt{h_1/2}} - \dfrac{\left(\varepsilon_1^2 + \varepsilon_2^2 \right)}{ \lambda} \nonumber \\ & -  \dfrac{U N}{2} \sum_{\alpha=R,B,G} {\boldsymbol{\phi}}_{\alpha} \cdot {\boldsymbol{\phi}}_{\alpha} \,.
\end{align}
We look for the value of $ | \boldsymbol{ \phi }  |$ which minimizes this energy. $\partial_{\phi^z} E_{\textrm{tot}} = 0$ gives
\begin{align} \label{solantif}
\dfrac{1}{U} = &\dfrac{1}{2N} \sum_{{\bf k}} \dfrac{1}{\sqrt{4\varepsilon_3^2({\bf k}) + U^2 | \boldsymbol{ \phi }  |^2 } } \nonumber \\ &\times \left( 1 + \dfrac{4\varepsilon_1  \varepsilon_2 \varepsilon_3 \cos \phi}{\lambda \left( 4\varepsilon_3^2({\bf k}) + U^2 | \boldsymbol{ \phi }  |^2 \right)} \right).
\end{align}
The solution of the previous equation is given in Fig.~\ref{mago}, from numerical evaluation at $\lambda=30t$.

\begin{figure}[t!]
\centering{\includegraphics[width=0.95\columnwidth]{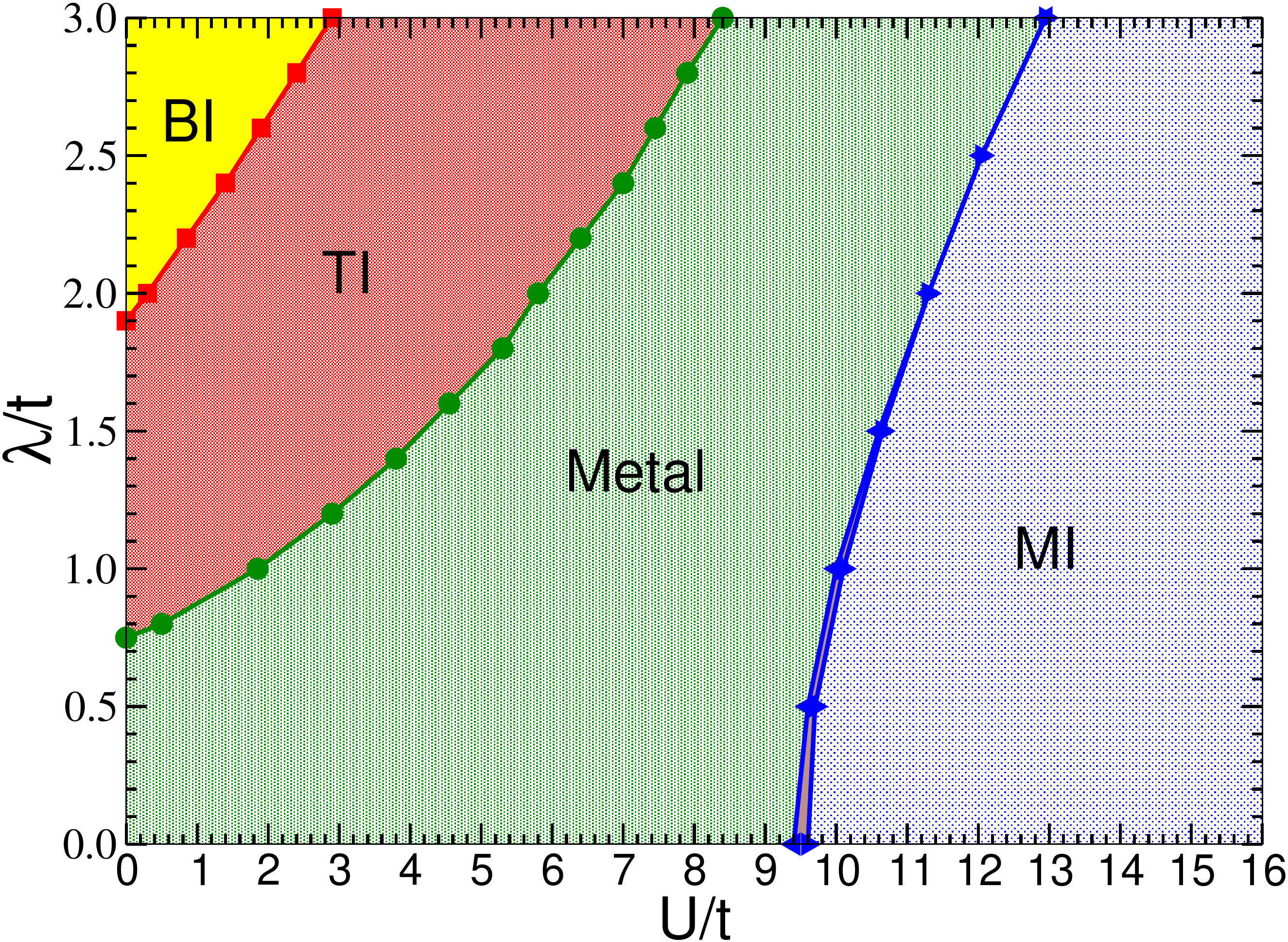}}
\caption{The paramagnetic phase diagram for the interacting system for half-filling and staggered onsite energies. $\phi=\pi/2$ and $\gamma=0.05$, $\lambda_{R,1}=\lambda_{B,1}=\lambda_{G,1}=-\lambda_{R,2}=-\lambda_{B,2}=-\lambda_{G,2}=\lambda$. The temperature is $T=0.1t$.
We obtain four distinguishable phases: band insulator (BI) [yellow region],  topological insulator (TI) [red region],  metallic phase [green region], and Mott insulator (MI) [blue region]. The phase transition between the BI and the TI (red curve with squares), as well as the one between the TI and the metallic phase (green curve with circles) are of second order. In contrast, the phase transition between the metallic phase and the MI (blue curves with triangles) is the first order. We obtain a hysteresis region (brown area between blue curves). 
}
\label{Fig:Phase-Diagram_Staggered_MI}
\end{figure}

\begin{figure}[t!]
\centering{\includegraphics[width=0.95\columnwidth]{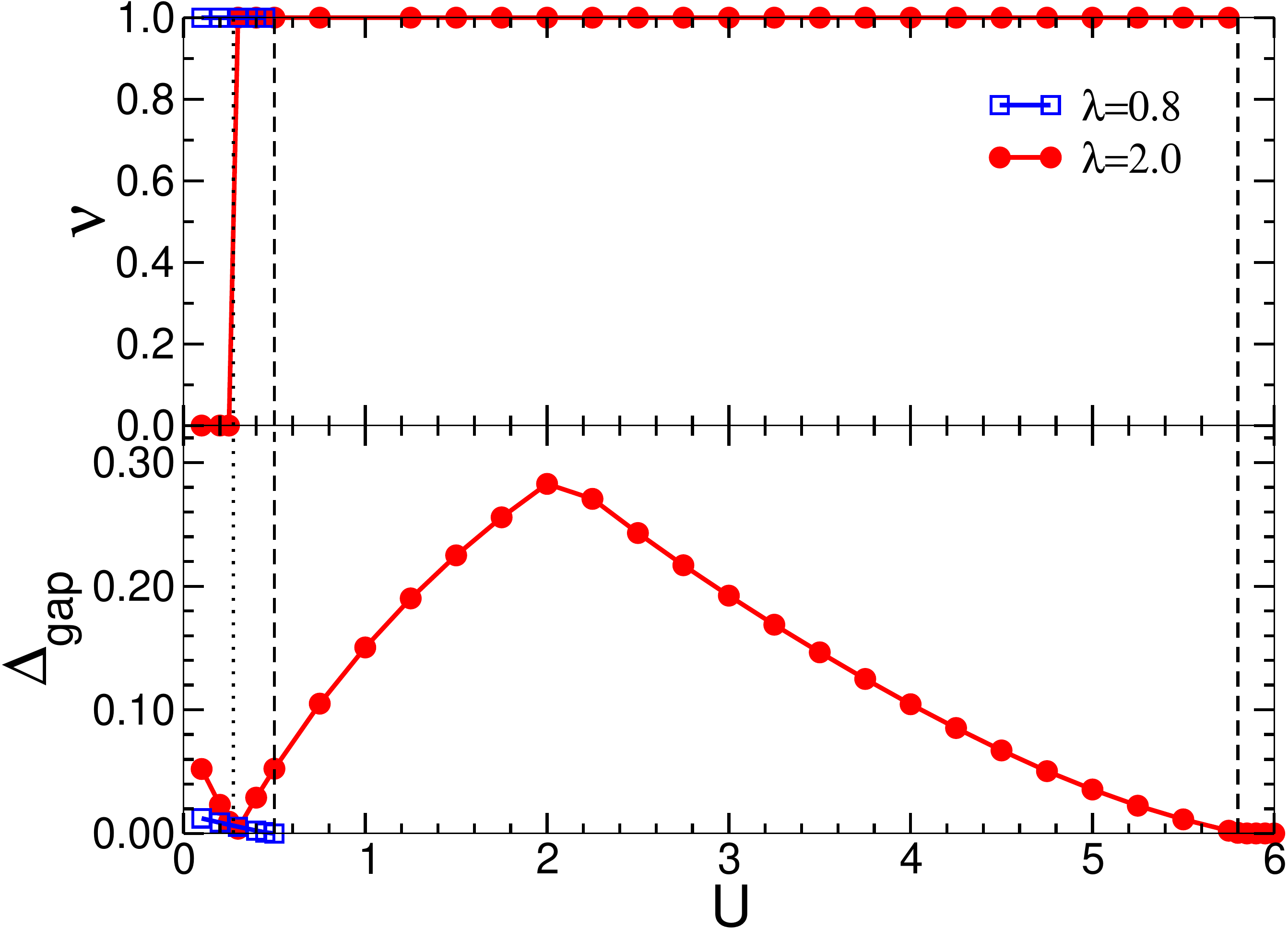}}
\caption{The $\mathbb{Z}_2$ number $\nu$ (upper panel) and gap $\Delta_{\rm gap}$ (lower panel) as a function of local Hubbard interaction $U$ for different values of onsite energies. 
Vertical dotted lines correspond to phase transition between the band and the topological insulators, while vertical dashed lines correspond transition between the topological phase and the metallic phase.  
Other parameters same as Fig. \ref{Fig:Phase-Diagram_Staggered_MI}.
}
\label{Fig:Gap_and_nu_Stgr}
\end{figure}

\begin{figure*}[t!]
\begin{center}
\subfigure[$\lambda=1.0t$]{
\label{Fig:M_MI_transition_Str_l0}
\includegraphics[width=0.3\textwidth]{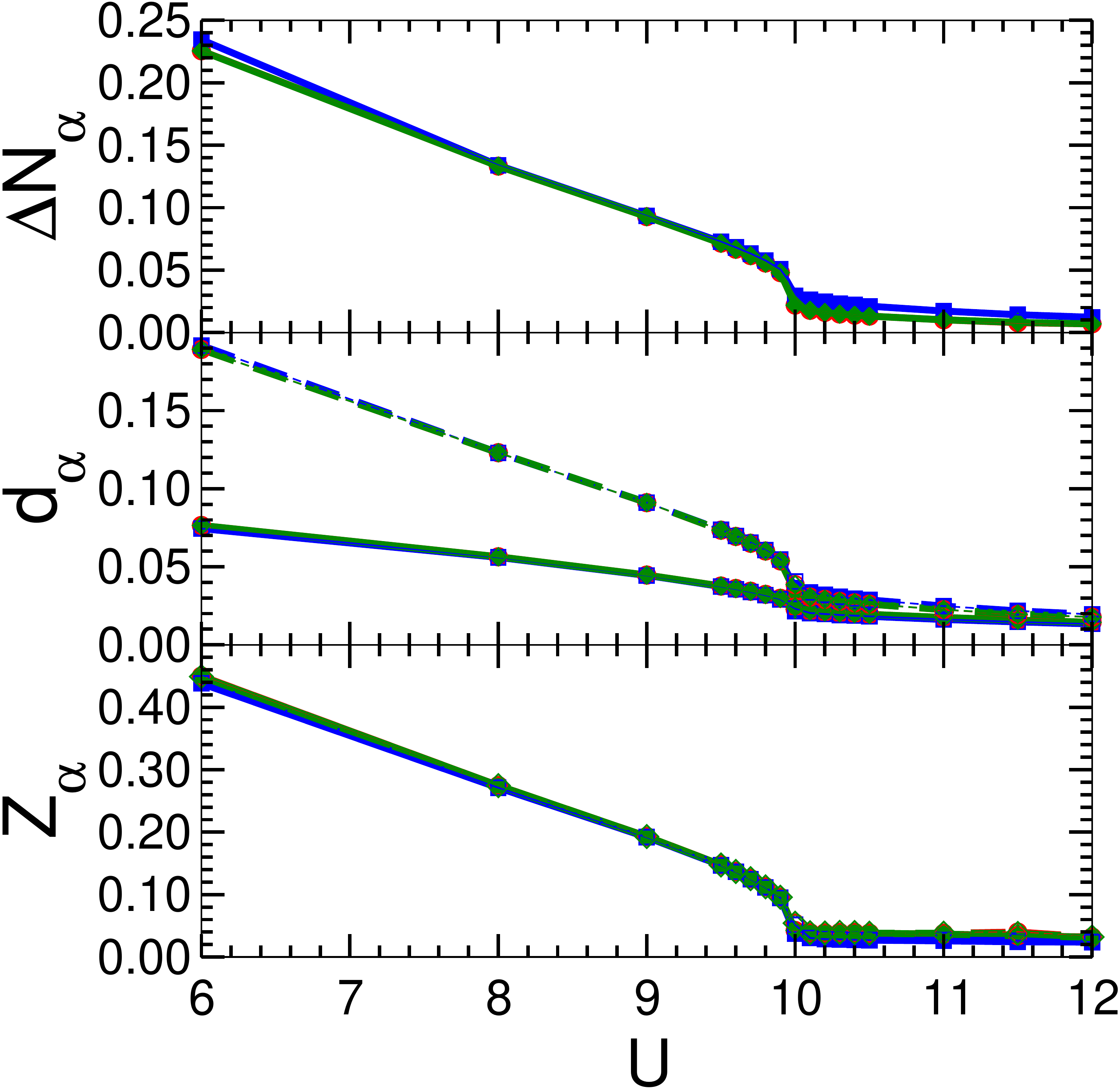} 
}
\hspace{0.005\textwidth} 
\subfigure[$\lambda=2.0t$]{
\label{Fig:M_MI_transition_Str_l0.5}
\includegraphics[width=0.3\textwidth]{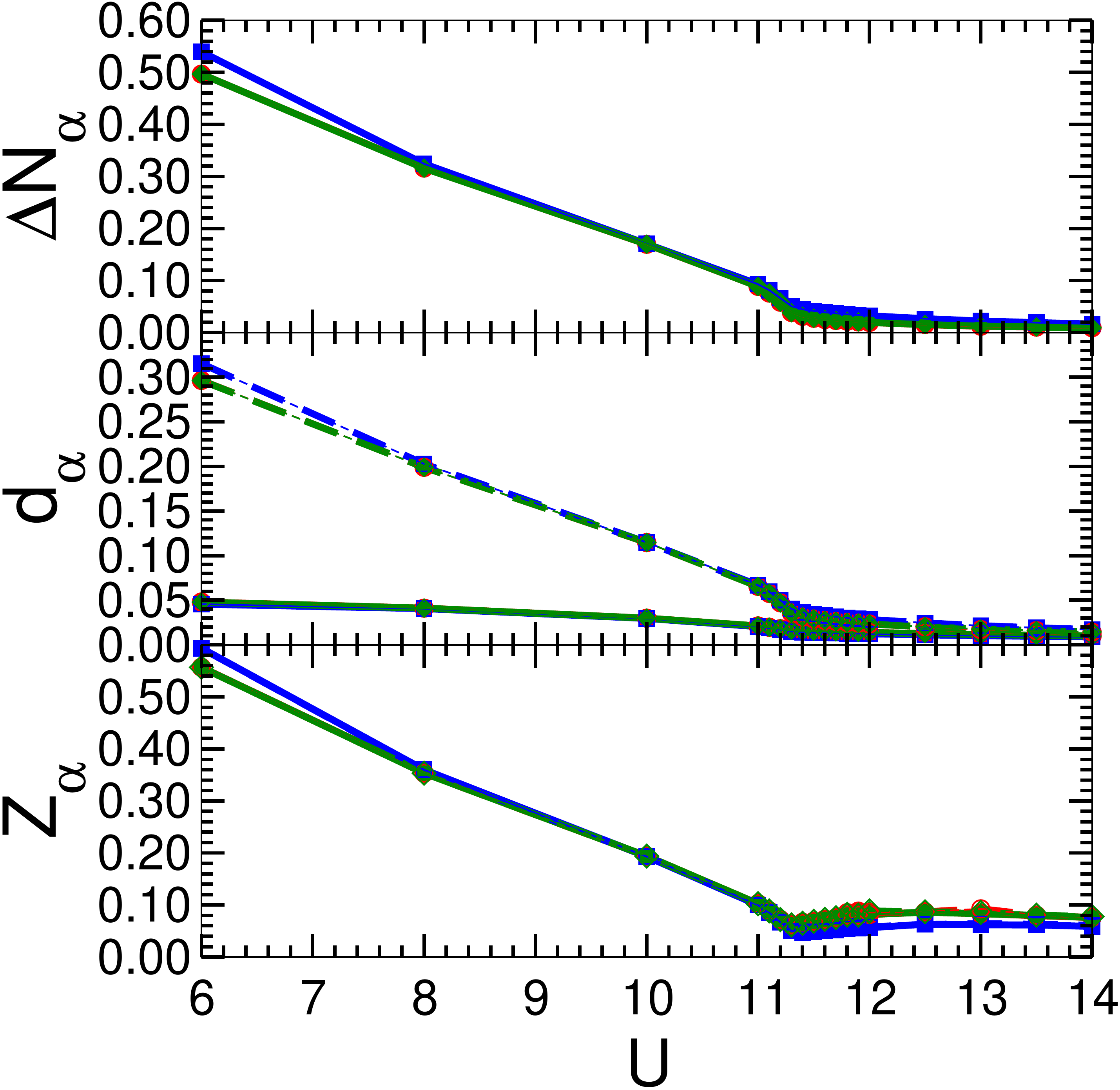} 
}
\hspace{0.005\textwidth} 
\subfigure[$\lambda=3.0t$]{
\label{Fig:M_MI_transition_Str_l1.0}
\includegraphics[width=0.3\textwidth]{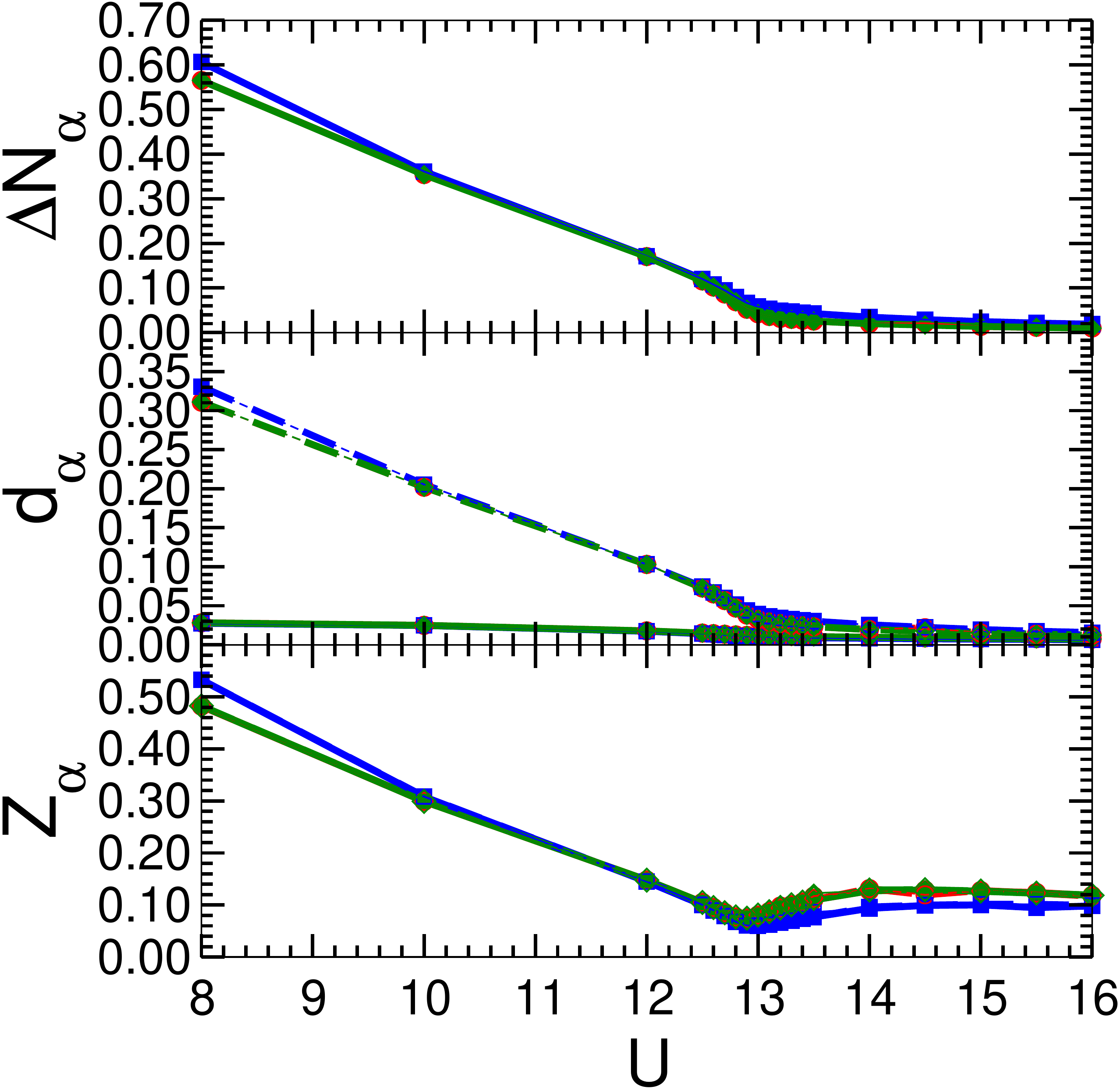} 
}
\end{center}
\caption{Double occupancy $d_\alpha$, quasi-particle weight $Z_\alpha$, and filling difference $\Delta N_\alpha=|N_{\alpha,2}-N_{\alpha,1}|$ between the two $\alpha$ sites in the model unit cell  as a function of Hubbard interaction $U$ for different values of onsite energy $\lambda$. Line colors (due to equal values not well visible) correspond to $R$ (red), $B$ (blue), and $G$ (green) sublattice sites.  The solid (dashed) lines are for $\kappa=1$ ($\kappa=2$), i.e. for first (second) $\alpha$-th site in the model unit cell. 
To detect hysteresis we start the DMFT iterations from different metallic (M) and insulating (I) initial self-energies corresponding to thin and thick curves, respectively. Other parameters same as Fig. \ref{Fig:Phase-Diagram_Staggered_MI}.}
\label{Fig:M_MI_transition_Str}
\end{figure*}

\begin{figure}[t]
\centering{\includegraphics[width=0.95\columnwidth]{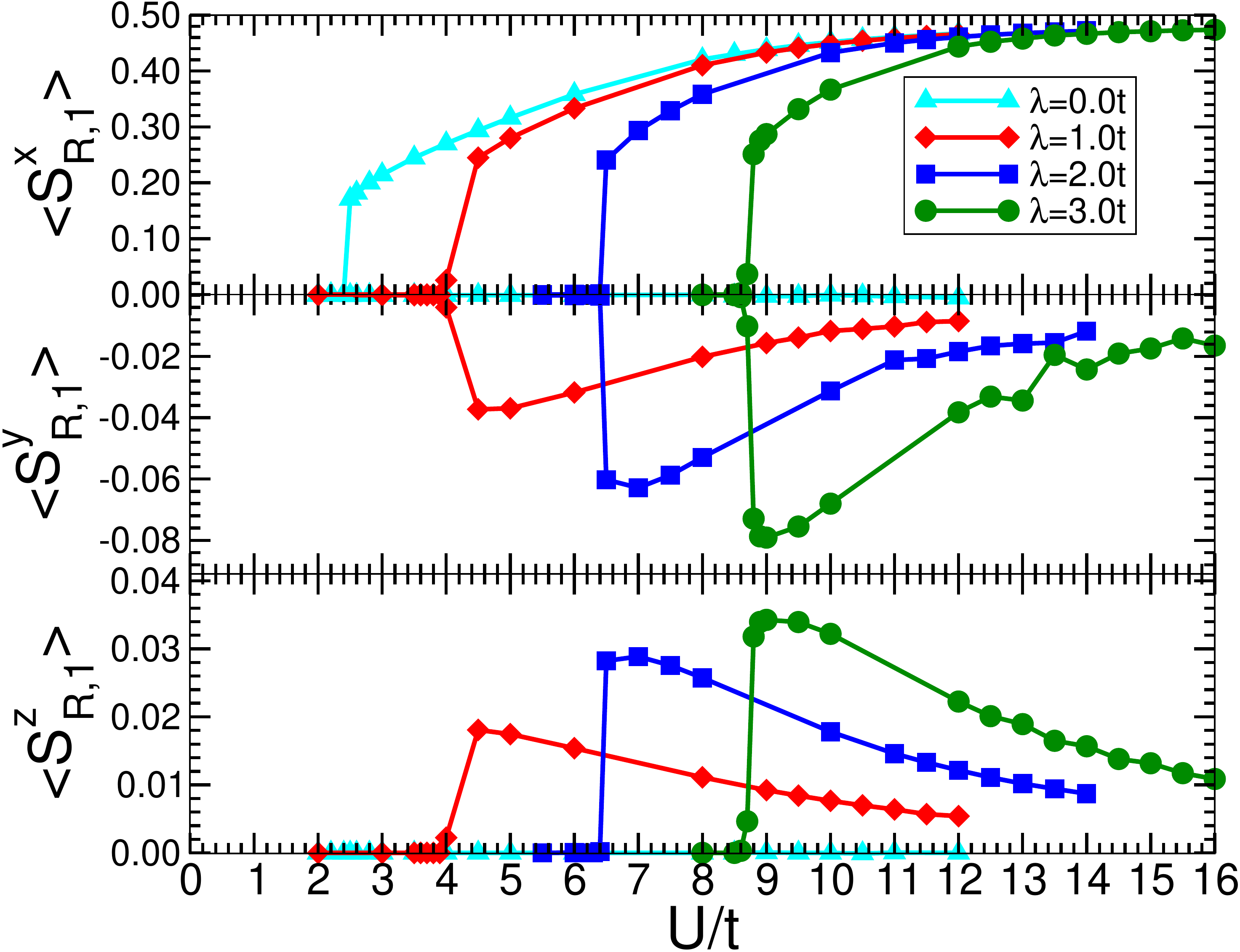}}
\caption{$\langle S_{R,1}^{x,y,z}\rangle$ as a function of local Hubbard interaction $U$ for different values of onsite energies. Other parameters same as Fig. \ref{Fig:Phase-Diagram_Staggered_MI}.
}
\label{Fig:SxSySz_vs_U_Stgr}
\end{figure}

\begin{figure}[t!]
\centering{\includegraphics[width=0.95\columnwidth]{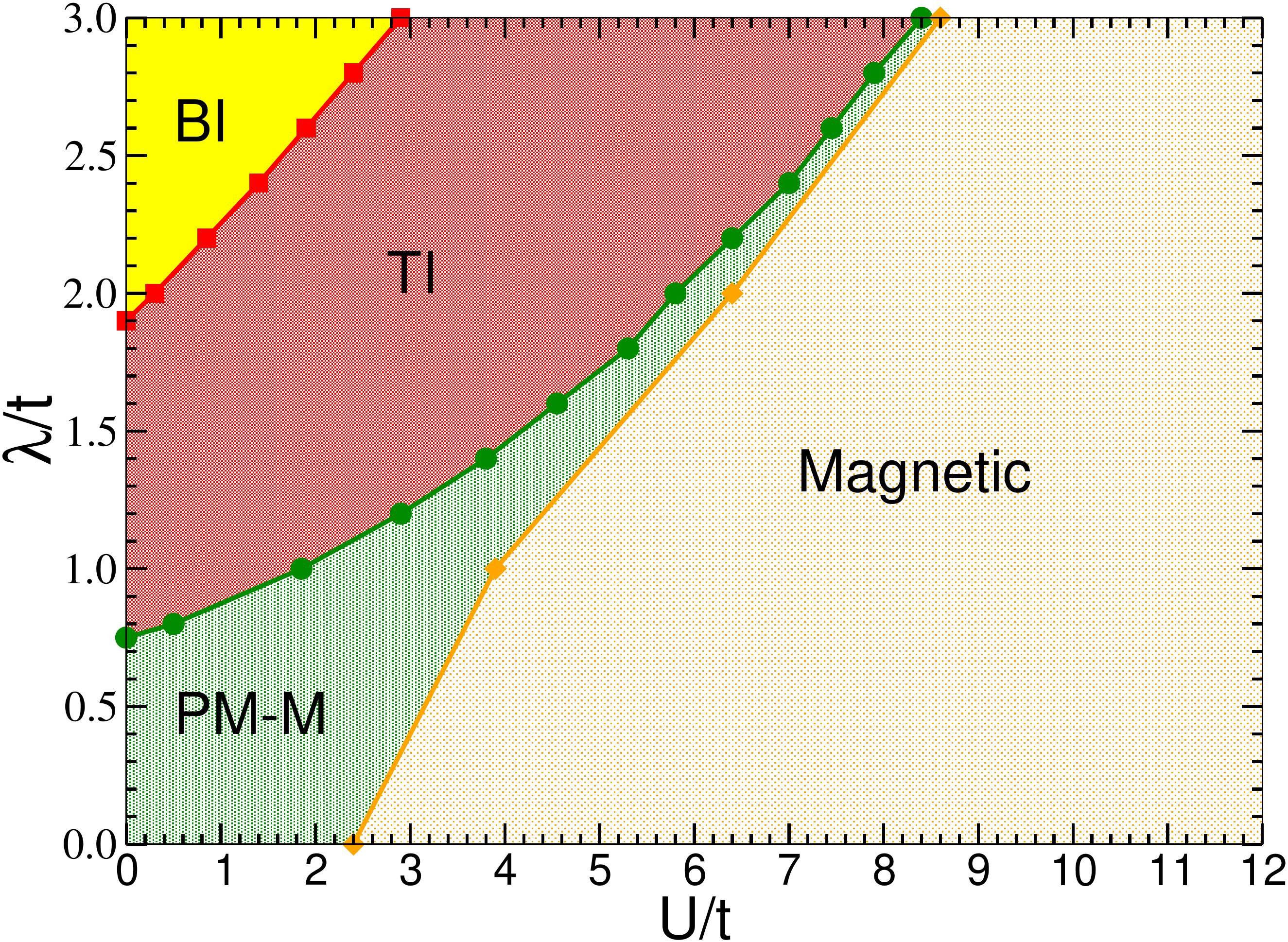}}
\caption{Magnetic phase diagram for the interacting system at half-filling and with staggered onsite energies. $\phi=\pi/2$ and $\gamma=0.05$, $\lambda_{R,1}=\lambda_{B,1}=\lambda_{G,1}=-\lambda_{R,2}=-\lambda_{B,2}=-\lambda_{G,2}=\lambda$. The temperature is $T=0.1t$.
We obtain four distinguishable phases: band insulator (BI)  [yellow region],  topological insulator (TI) [red region], paramagnetic metallic phase (PM-M) [green region],  and magnetic phase [orange region]. All phase transitions are of second order. 
}
\label{Fig:Phase-Diagram_Staggered_Mag}
\end{figure}

\section{Results: staggered potential}
\label{section_staggered_potential}

In this section we study the system with staggered potential. As it was mentioned earlier this corresponds to the onsite potential ${V_{\alpha,{\bf r}}=\lambda}$ for ${\bf r}=2n_1{\bf e}_1+n_2{\bf e}_2$ and  
${V_{\alpha,{\bf r}}=-\lambda}$ for ${\bf r}=(2n_1+1){\bf e}_1+n_2{\bf e}_2$. 
We again consider the flux $\phi=\pi/2$, but unlike the previous section, we consider finite spin-orbit coupling $\gamma=0.05$.  Here we study a half-filled system. 
We have shown in Ref.~\onlinecite{ti.le.21} that in the non-interacting limit for this choice of parameters we have three different phases: 
the metallic phase for $\lambda<0.75t$, the topological insulator for $0.75t<\lambda<1.9t$, and the band insulator for $\lambda>1.9t$. Our goal here is to study the effect of the Hubbard interaction. 
Our calculations are again performed by R-DMFT. As system is particle-hole symmetric to obtain half-filling we fix the  chemical potential as $\mu=U/2$. In the calculations presented below, we again consider the system size $N_1=N_2=20$ with periodic boundary conditions. Due to the symmetry of the model, we consider $6$ distinguishable self-energies 
$\Sigma_{R,1,\sigma}(i\omega_n)$, $\Sigma_{B,1,\sigma}(i\omega_n)$,  $\Sigma_{G,1,\sigma}(i\omega_n)$, $\Sigma_{R,2,\sigma}(i\omega_n)$, $\Sigma_{B,2,\sigma}(i\omega_n)$, and $\Sigma_{G,2,\sigma}(i\omega_n)$.

Our results are summarized in the phase diagram in Fig.~\ref{Fig:Phase-Diagram_Staggered_MI}. We obtain four different phases: the topological insulator (TI), the band insulator (BI), the Mott insulator (MI), and the metallic phase. 
To determine the transitions between different phases, we first calculated the $\mathbb{Z}_2$ number and the gap $\Delta_{\rm gap}$ as a function of the Hubbard interaction $U$ for different values of the onsite energies $\lambda$ based on the topological Hamiltonian.  Our results are shown in Fig.~\ref{Fig:Gap_and_nu_Stgr}. 
We find that for $\lambda<0.75t$ the gap $\Delta_{\rm gap}=0$ and consequently the system is in the metallic phase. For $0.75t<\lambda<1.9t$, for weak interactions, the gap $\Delta_{\rm gap}$ is finite and the $\mathbb{Z}_2$ number $\nu=1$, which implies that the system is in the TI phase. As the interaction strength increases, the size of the gap $\Delta_{\rm gap}$ decreases and at a critical value $U=U_{c}^{\rm TI-M}(\lambda)$, the gap closes and remains closed even after further increasing the interaction strength $U$. This indicates the transition to the metallic phase (blue curve in Fig.~\ref{Fig:Gap_and_nu_Stgr}). For $\lambda>1.9t$ at weak interactions, the gap $\Delta_{\rm gap}$ is finite and the $\mathbb{Z}_2$ number $\nu=0$, suggesting that the system is in the BI phase. As the interaction strength $U$ increases, the size of the gap $\Delta_{\rm gap}$ decreases, for a certain critical value $U=U_{c}^{\rm BI-TI}(\lambda)$ the gap closes and after further increase of $U$ the gap opens again, but now the $\mathbb{Z}_2$ number is $\nu=1$, indicating that the system is in the TI (red curve in Fig.~\ref{Fig:Gap_and_nu_Stgr}). After further increasing the interaction $U$, the size of the gap reaches a maximum and then decreases again. At a critical value  $U=U_{c}^{\rm TI-M}(\lambda)$, the gap closes and remains closed even after further increasing the interaction strength $U$. This indicates a transition to the metallic phase (red curve in Fig.~\ref{Fig:Gap_and_nu_Stgr}).

Furthermore, we investigate the possible transition to the Mott insulator phase. To detect the transition from the metallic to the Mott insulator phase, we calculate the double occupancy $d_{\alpha,\kappa}$, the quasi-particle weight $Z_{\sigma,\alpha,\kappa}$, and the filling difference between the two $\alpha$-sites within the unit cell of the model $\Delta N_{\alpha}=|N_{\alpha,2}-N_{\alpha,1}|$.  
Here $\kappa=1,2$ numbers the $\alpha$-site in the unit cell.

Here we perform calculations in the paramagnetic phase, thus $Z_{\alpha,\kappa,\uparrow}=Z_{\alpha,\kappa,\downarrow}$.  Our results are shown in Fig. \ref{Fig:M_MI_transition_Str}. For $\lambda=0$, the number of particles in the first and the second $\alpha$-sites are equal, while for finite onsite energies $\Delta N_\alpha >0$. 
We obtain that the double occupancy $d_{\alpha,\kappa}$, the quasi-particle weight $Z_{\sigma,\alpha,\kappa}$, and the filling difference between two $\alpha$-sites within the unit cell $\Delta N_{\alpha}$ (for $\lambda\neq0$) decrease with the increase of the Hubbard interaction $U$. They show a kink-like behavior for $U=U_{c}^{\rm M-MI}(\lambda)$, suggesting the transition to the Mott insulator phase. In the Mott insulator, all these three quantities are much smaller than one.

To further investigate the phase transition, we perform calculations with different initial self-energies: a metallic and a Mott-insulating one. For weak onsite energies ${\lambda \lessapprox 1.5t}$ near the phase transition, we obtain two different solutions, the metallic and the Mott-insulator solution, depending on whether we start the DMFT iterations from a metallic or a Mott-insulating  self-energy. This indicates the existence of hysteresis. It means that the transition is first order. For intermediate and large values of the onsite energies, we cannot detect any hysteresis. This means that the hysteresis is either beyond our numerical accuracy or that the character of the transition changes from first order to second order.

Finally, we again remove the paramagnetic constraint and study the magnetic properties of the system. For this purpose, we study $\langle S_{\alpha,\kappa}^{x,y,z}\rangle$ for different sublattice sites as a function of the Hubbard interaction $U$ for different onsite energies $\lambda$ (see Fig.~\ref{Fig:SxSySz_vs_U_Stgr}). 
For weak interactions, the system is in the paramagnetic phase and $\langle S_{\alpha,\kappa}^{x}\rangle = \langle S_{\alpha,\kappa}^{y}\rangle = \langle S_{\alpha,\kappa}^{z}\rangle =0$. 
So all the results obtained are the same as shown above and we recover the same phases.
With increasing interaction strength for $U=U_c^M$, the transition to the magnetic phase occurs. 
We obtain that
\begin{align*}
&S_{R,1}^x=S_{R,2}^x = - S_{G,1}^x = - S_{G,2}^x
 \simeq -S_{B,1}^x= -S_{B,2}^x
\\
&S_{R,2}^y=-S_{R,1}^y =  S_{G,2}^y = - S_{G,1}^y
\ll 1 \,,\,\,\, 
S_{B,1}^y=-S_{B,2}^y=0\\
&S_{R,1}^z=-S_{R,2}^z = - S_{G,1}^z =  S_{G,2}^z 
\ll 1\,, \,\,\,
S_{B,1}^z=-S_{B,2}^z=0 \,.
\end{align*}
Here we note that for $\lambda=0$ only the $x$ component of the spin is different from zero, i.e. $\langle S_{\alpha,\kappa}^{y}\rangle \simeq \langle S_{\alpha,\kappa}^{z}\rangle \simeq  0$. Here $\alpha=R,B,G$ and $\kappa=1,2$.

We observe that for a given onsite energy $\lambda$, the relation $U_c^M(\lambda) \geq U_{c}^{\rm TI-M}(\lambda)$ holds and as the onsite energy $\lambda$ increases, these two critical values of the interaction converge (see also Fig.~\ref{Fig:Phase-Diagram_Staggered_MI}).

\section{Conclusions}
\label{Conclusions}

In this work, we have studied the Hubbard model with time-reversal invariant flux and spin-orbit coupling and position-dependent onsite energies on the kagome lattice, which is a non-Bravais lattice and has three sites per unit cell. 
To investigate it, we applied R-DMFT, a powerful method for studying strongly correlated systems in two and more dimensions. R-DMFT allows us to study the transition to the Mott insulator phase or to the magnetic phase. In addition, to study the topological transition, we used the topological Hamiltonian method. To gain more insight into the phase transition to the magnetic phase, we have applied analytical methods based on perturbation theory for strong interactions and large onsite energies, and on stochastic mean-field theory.

We have investigated two different setups. First, we considered the case where $\phi=\pi/2$, $\gamma=0$ and the onsite energies are applied only to $R$ sublattice sites. In this case, we consider $2/3$ filling. 
First, we performed paramagnetic calculations. For weak and intermediate interactions, we obtained band and topological insulators. 
For large onsite energies, the system can be described by an effective model on a half-filled square lattice\cite{ti.le.21}. We have shown that as the interaction increases, the transition to the Mott insulator phase occurs.  
We also investigated the magnetic properties of the system. Our R-DMFT calculations show that at weak interactions the system is in the paramagnetic phase, while as the interaction strength increases there is a transition to the antiferromagnetic phase. 
We obtain a hysteresis region were paramagnetic and magnetic solutions coexist. 
To gain more insight into this phase transition, we have also used analytical methods, as mentioned above. The results obtained are in full agreement with our R-DMFT calculations. 
Using these analytical methods, in addition to $\phi=\pi/2$, we also perform calculations for $\phi <\pi/2$.  Also for the latter case we obtain an antiferromagnetic phase.

Another setup we consider is the staggered potential at half-filling. Within the unit cell of the kagome lattice, the onsite energies are equal, but they oscillate along the ${\bf e}_1$ direction. 
We studied the system for $\phi=\pi/2$ and $\gamma=0.05$ using R-DMFT. 
First, we again performed paramagnetic calculations and obtained four different phases: band, topological and Mott insulators and a metallic phase. 
For strong interactions, we observed the transition to the Mott insulator phase. The critical value of the interaction increases with the increase of the onsite energies, in contrast to the case where onsite energies are applied only to $R$ sublattice sites. 
Using R-DMFT, we have also studied the magnetic properties of the system. We observed that the system is paramagnetic at weak interactions, while with increasing interaction strength the transition to the magnetic phase occurs. The magnetization along the $x$ direction dominates, although the magnetization along the $y$ and $z$ directions is finite.

In summary, we have studied the effects of the interaction on the topological properties of the system. We also investigate the transition to the Mott insulator phase, as well as the magnetic properties of the system. 
Our model can be realized in the experiments with ultracold atoms. Therefore, it would be interesting to compare the theoretical predictions presented in this work with future experimental results.

\begin{acknowledgments}
This work was supported by the Deutsche Forschungsgemeinschaft (DFG, German Research Foundation) under Project No.~277974659 via Research Unit FOR 2414. 
This work was also supported by the DFG via the high-performance computing center {\it Center for Scientific Computing (CSC)}.  
The research on the topological kagome lattice is also funded by ANR BOCA (KLH) for which JL is also supported for his PhD. The authors thank Bernhard Irsigler and Maarten Grothus for helpful discussions. 
\end{acknowledgments}



%

\end{document}